\documentclass[10pt,journal]{IEEEtran}
\IEEEoverridecommandlockouts
\usepackage[dvips]{graphicx}
\usepackage[colorlinks,urlcolor=blue,linkcolor=blue,citecolor=blue]{hyperref}
\usepackage{color,array}
\usepackage{xspace}
\usepackage{amsmath}
\usepackage{caption}
\usepackage{subcaption}
\usepackage{placeins}
\usepackage{balance}
\usepackage{cite}
\usepackage{bm}
\usepackage{amsmath}
\usepackage{array}
\usepackage{setspace} 
\usepackage[nameinlink]{cleveref}
\usepackage{accents}
\usepackage{mathtools}
\usepackage{algpseudocode}
\usepackage[dvipsnames]{xcolor}
\usepackage{amsthm} 
\usepackage{multirow}
\usepackage[linesnumbered,ruled,vlined]{algorithm2e}
\usepackage{tabularx}
\SetKwRepeat{Do}{do}{while}
\usepackage{color}
\usepackage{epstopdf}
\usepackage{endnotes}
\usepackage{framed}
\usepackage{float}

\usepackage{cool} 
\usepackage{enumerate} 
\usepackage{pifont}
\Crefname{figure}{Fig.}{Figs.}

\usepackage{mathrsfs}
\usepackage[justification=centering]{caption}
\allowdisplaybreaks[4]

\newcolumntype{P}[1]{>{\centering\arraybackslash}p{#1}}

\DeclareUnicodeCharacter{2212}{-}

\usepackage{changes}
\usepackage{lipsum}
\definechangesauthor[name={Moqbel1}, color=red]{Revised/Added}
\definechangesauthor[name={Moqbel2}, color=red]{Deleted}
\definechangesauthor[name={Moqbel3}, color=red] {moved to appendix E}

\usepackage{color}

\def\BibTeX{{\rm B\kern-.05em{\sc i\kern-.025em b}\kern-.08em
        T\kern-.1667em\lower.7ex\hbox{E}\kern-.125emX}}
\begin{document}

 \title{Optimized Federated Multitask Learning in Mobile Edge Networks: A Hybrid Client Selection and Model Aggregation Approach}
\author{Moqbel Hamood,
      Abdullatif Albaseer,~\IEEEmembership{Member,~IEEE,}
        Mohamed Abdallah,~\IEEEmembership{Senior Member,~IEEE,}
        Ala Al-Fuqaha,~\IEEEmembership{Senior Member,~IEEE,}
        Amr Mohamed,~\IEEEmembership{Senior Member,~IEEE}
    \thanks{Copyright (c) 20xx IEEE. Personal use of this material is permitted. However, permission to use this material for any other purposes must be obtained from the IEEE by sending a request to pubs-permissions@ieee.org.}
    \thanks{Moqbel Hamood, Abdullatif Albaseer, Mohamed Abdallah and~Ala Al-Fuqaha and are with the Division of Information and Computing Technology, College of Science and Engineering, Hamad Bin Khalifa University, Doha, Qatar (e-mail:\{moha19838,aalbaseer, moabdallah, aalfuqaha\}@hbku.edu.qa).}
    \thanks{Amr Mohamed is with the Division of Computer Science and Engineering, College of Engineering, Qatar University, Doha, Qatar (e-mail:\{amrm\}@qu.edu.qa).

}
}
\maketitle
\begin{abstract}
Clustered federated multitask learning is introduced as an effective strategy, specifically designed to tackle statistical challenges, including non-independent and identically distributed data among clients. This is achieved by clustering clients based on a similarity in their data distributions and assigning a specialized model for each cluster. However, this approach encounters complexities when applied in hierarchical wireless networks due to hierarchical two-level model aggregation and resource-based client selection. This results in a slower convergence rate and deprives clients of the best-fit specialized model for those having similar data distribution from different mobile edge networks. To this end, we propose a framework comprising two-phase client selection and two-level model aggregation schemes designed for IoT devices and intelligent vehicles. For client selection, the cloud ensures fairness among all clients by providing them with equal priority to participate in training, contributing to more accurate clustering.  Once a particular cluster reaches a stopping point, the related edge server performs two client selection methods separately, greedy and round-robin. The greedy algorithm prioritizes clients with less latency and superior resources, while the round-robin algorithm allows for cyclic participation in training. The cloud then executes model aggregation in two distinct ways: one based on reaching pre-determined aggregation rounds (round-based model aggregation) and the other by performing at least one split at the edge servers (split-based model aggregation). We conduct extensive experiments to evaluate our proposed approach. Our results show that the proposed algorithms significantly improve the convergence rate, minimize training time, and reduce energy consumption by up to 60\%  while providing every client with a specialized model specifically attuned to its data distribution.
\end{abstract}
\begin{IEEEkeywords}
Clustered Federated learning (CFL), Hierarchical mobile wireless networks, Model aggregation, Client selection, Resource allocation.
\end{IEEEkeywords}
\section{Introduction}
\IEEEPARstart{T}he advent of the Internet of Things (IoT) technology, strengthened by the fifth-generation (5G) networks, has engendered a range of innovative applications such as autonomous driving and traffic prediction in intelligent vehicles, with each producing heterogeneous data in large volumes~\cite{sun2019application,shu2021driving,thai2022machine,faisal2022machine}. The surge in IoT applications and the resultant increase in generated data present unprecedented opportunities for cross-application knowledge sharing, particularly within the context of hierarchical wireless networks (HWNs), enabling different IoT networks to collaborate with each other. Specifically, data communication operates on two levels – first via the edge servers and subsequently through the cloud – facilitating efficient and robust data exchange.

However, the vast amount of data produced by IoT edge devices, including those used in applications such as speech recognition~\cite{krizhevsky2017imagenet,9933029}, computer vision~\cite{jia2014caffe}, and traffic prediction for autonomous vehicles poses significant challenges.  The traditional practice of offloading this data to the cloud for machine learning (ML) analysis proves to be resource-intensive and costly, owing to network constraints and the intrinsic heterogeneity of the data~\cite{rodrigues2019machine, li2020federated}. Furthermore, this methodology of centralized data processing presents substantial privacy concerns, as sensitive information is often embedded within the offloaded data.

In response to these challenges, hierarchical federated learning (HFL) emerged as a promising solution~\cite{konevcny2016federated,abad2020hierarchical,chen2022federated,albaseer2021fine}, where data is kept at its source, thereby preserving privacy. In contrast to the conventional FL, which consists of a two-component structure within edge networks  (i.e., the central server and the edge devices),  the HFL expands this configuration by incorporating more aggregation layers, forming a hierarchical structure. More specifically, the HFL is implemented in HWNs consisting of three main layers: the cloud server at the top, the edge servers in the middle, and the edge devices at the bottom. In this approach, client devices such as intelligent vehicles participate in the training process, sending only model parameters, such as weights and biases, to the edge servers~\cite{10057044,ji2023efficiency}. These edge servers aggregate the received updates, create edge models, and then upload them to the cloud. The cloud then collects these edge models to update its global model, which is sent back to the clients in different edge networks for further training or future use. This iterative process continues until the global model converges, offering a promising framework for optimizing resource utilization and reducing latency while preserving privacy. Despite these advantages, the HFL faces several critical challenges, including system and statistical heterogeneity~\cite{niknam2020federated,zhang2022federated}.

System heterogeneity arises from the different capabilities of the clients due to variations in their hardware, storage, network connections, and power supplies~\cite{wang2022accelerating,albaseer2022semi,wang2023towards}. On the other hand, statistical heterogeneity refers to the data being diverse, unevenly distributed, and non-independent and identically distributed (non-IID)~\cite{van2009multi,khan2021federated,lee2022data}. Communication efficiency also presents a critical challenge in HWNs, as these networks usually comprise numerous edge devices, including millions of smartphones, intelligent vehicles, and IoT devices. These devices face resource constraints such as bandwidth and energy, which can prevent effective communication within HWNs. Consequently, these challenges restrict the number of devices participating in training rounds~\cite{rahman2021challenges,10193846,chen2021distributed}. Further, the challenges extend beyond system and data heterogeneities, involving problems related to resource consumption directly associated with transmitting models to the edge server-connected cloud every round. This highlights the crucial role of appropriate client selection and model aggregation techniques that increase the convergence speed and develop an unbiased model that fits incongruent data distributions while reducing resource consumption.

As a result, a promising approach proposed to address the above challenges, especially the unbalanced and non-IID data, is the clustered federated multitask learning (CFL) paradigm~\cite{sattler2020clustered,albaseer2021client,ghosh2022efficient}.  Unlike the standard HFL assumption that develops a shared model across clients despite variations in their data distribution, the CFL has the ability to group clients with similar data distributions into clusters and create tailored models, accurately accommodating the data patterns within each cluster. Note that the HFL assumption is violated in practical applications for two reasons~ \cite{sattler2020clustered}: first, the edge and cloud servers cannot check data owned by clients for privacy concerns, and second, the clients' hardware capabilities inherently limit the model's complexity. Thus, the CFL is only applied after the HFL solution reaches a stationary point (i.e., no further improvement on the HFL model using the current data) without making any adjustments to the HFL algorithm itself. Notably, the clustering process in CFL is applied to the HFL in two phases, enabling implicit knowledge sharing across the edge networks and enhancing the overall learning process. The first clustering occurs at the edge layer, where clients use HFL to train their local models and upload them to their associated edge servers. Each edge server then performs cosine similarity computations on the clients' model gradients to decide if clustering is necessary. When partitioning is needed, each cluster develops a specialized model tailored to its data distributions. The second clustering phase takes place at the cloud layer, where edge servers upload their specialized models to the cloud. The cloud then performs similarity computations among these models to capture more patterns across the HWNs' edge networks, effectively facilitating knowledge exchange. It is worth noting that each cluster, equipped with a specialized model, is dedicated to a specific learning task tailored to that cluster's data distribution or application requirements.

Several works have mainly focused on applying the CFL to edge networks, effectively tackling statistical challenges such as non-IID data distribution. For instance, the works in~\cite{duan2021flexible,agrawal2021genetic,gong2022adaptive,li2022towards} proposed approaches to address unbalanced and non-IID data distributions through collaborative training among clients within the same edge network.  Additionally, the authors  in~\cite{albaseer2021client,albaseer2023fair}  introduced fairness-aware approaches, enhancing learning performance and the convergence rate by ensuring fair client selection and scheduling. They also addressed the challenges associated with struggling clients during training by reusing bandwidth in the edge wireless network. Nevertheless, none of these studies~\cite {duan2021flexible,agrawal2021genetic,gong2022adaptive,li2022towards,albaseer2021client,albaseer2023fair} have considered collaboration among different edge networks to achieve better collaborative models. The significant challenges for these approaches stem from their inability to capture the data patterns across different edge networks, mainly because these approaches are typically implemented within a single edge network. To efficiently identify these patterns, a two-level model aggregation process is essential: one level at the edge servers to catch data patterns from diverse edge devices and a subsequent level at the cloud to capture the patterns across various mobile edge networks. This process necessitates a dual computation of similarity, executed in parallel with each model aggregation: initially at the edge servers to group clients with similar data distributions and subsequently in the cloud to provide collaborative learning across different mobile edge networks. Consequently, the issue of inter-edge network collaboration in CFL presents a critical challenge that is yet to be investigated; thus, demanding more research and exploration.

In response to these observations, we propose a new framework integrating two-phase client selection and two-level model aggregation schemes in HWNs. The framework has powerful capabilities to minimize latency and optimize resource consumption while tackling these networks' unbalanced and non-IID data distribution. It also aims to provide collaborative training even among different mobile edge networks in HWNs to catch diverse patterns, achieving a better collaborative global model, all while considering both resource and deadline constraints by the cloud and edge servers to avoid lengthy delays before starting a new global round. To this end, we need dual model aggregation alongside two-level similarity computation. Firstly, edge servers aggregate local models and compute similarities between client gradients for clustering. Next, specialized models from different edge servers are collected and analyzed in the cloud to capture diverse patterns, enhancing collaborative learning. It is worth mentioning that the proposed approach is designed with a broad range of applications in mind, focusing more on IoT devices and intelligent vehicles as primary examples due to their urgent need for real-time data processing and efficient distributed learning. However, the strategies supporting the proposed approach can generally be applied for various applications, addressing critical challenges such as data and system heterogeneities and network resource constraints. The key contributions of this work are as follows.
\begin{itemize}
   \item Proposing a novel framework encompassing two-phase client selection methods and two-level model aggregation schemes to address diverse patterns across varying mobile edge networks, thereby enabling collaborative knowledge exchange through the creation of a shared global model. This structure also addresses the challenges of limited resources and increases communication capabilities in HWNs by minimizing training time and energy consumption while tackling the unbalanced and non-IID data distribution issue through efficient clustering techniques in a hierarchical manner.  Our framework enables fairness by ensuring all clients, regardless of their channel states and data sizes, have an equal opportunity to participate in the training process. Considering these factors, our proposed approach delivers highly customized models for clients within HWNs.

   \item Formulating a joint optimization problem for client selection and model aggregation to reduce training time as well as resource consumption while accelerating convergence speed. This considers the unbalanced and non-IID data distribution, device heterogeneity, and resource scarcity across various mobile edge networks within the HWN. Because of the NP-hardness of this problem, we propose a heuristics-based solution leveraging the heterogeneity of devices and the two-level model aggregation of client models in both edge and cloud layers.
   
   \item Conducting experimental evaluation using both FEMNIST and CIFAR-10 datasets, known for their unbalanced and non-IID data distributions. Experimental results verify that our proposed solution significantly reduces both training time and resource consumption and enhances convergence speed while obtaining satisfactory performance.
\end{itemize}
The remainder of the paper is organized as follows. We review related works in Section~\ref{related_work}. The system model, learning process, and models for localized computation and communication are presented in Section~\ref{Sys_model}. Section~\ref{problem_form} introduces the Problem formulation. The proposed approach is detailed in Section~\ref{proposed_sol}.
Section~\ref{results} delineates our experimental results and findings. Lastly,  Section~\ref{conclusion} concludes this work and provides pointers for future research directions.
\section{Related Work}
\label{related_work}
\textbf{FL in HWNs:} Studying FL in HWNs has drawn considerable interest from researchers due to its ability to optimize time and resource utilization~\cite{feng2021min, cui2022optimizing,abdellatif2022communication}. For example, the work in~\cite{luo2020hfel} proposed the scheduling and two-level model aggregation for FL in HWNs to reduce computation and communication costs by transmitting models to the cloud via edge servers. 
Also, the authors in~\cite{xu2021adaptive} suggested an adaptive HFL system that integrated edge aggregation interval control and resource allocation to minimize the combined sum of training loss and latency. 
The work in~\cite{feng2021min} proposed a min-max cost-optimal problem that inherently affects overall HFL performance. The authors aimed to achieve optimal results by factoring in participants' local accuracy, subcarrier assignment, transmission power allocation, and computational resource allocation.
Besides, the authors in~\cite{lim2021dynamic} proposed a hierarchical game-theoretic framework that accurately models the dynamic, self-organizing characteristics of edge association and resource allocation within HFL. Lastly, the researchers in~\cite{liu2022hierarchical} introduced a communication-efficient training algorithm for HFL. This approach, designed to accelerate the convergence rate, employed model quantization to decrease communication overhead during model uploading. However, model quantization can significantly affect model performance and convergence rate, especially when accuracy reduces. Despite their contributions, the previously mentioned studies either ignored the challenge of statistical heterogeneity or struggled with non-IID data, mainly when dealing with highly diverse data distributions.  Therefore, it can be inferred that the scheduling techniques in the literature for HFL cannot be directly applied to CFL.
\\
\textbf{CFL in Edge Networks:} CFL has recently emerged as a promising solution to tackle statistical challenges such as non-IID data distributions and restrictions in model complexity inherent to FL~\cite{ghosh2020efficient,luo2021energy,gong2022adaptive}. For instance, authors in~\cite{sattler2020clustered} introduced a CFL framework that clusters clients with similar data distributions for joint training, leveraging the geometric characteristics of the FL loss surface. This study provided a robust mathematical justification for the effectiveness of clustering in CFL.  Other studies have also proposed solutions to bandwidth limitations and energy consumption challenges in the CFL. For instance, the work in~\cite{duan2021flexible}  introduced a flexible CFL technique that groups clients based on their optimization direction similarities and ensured scalability by adding newcomer devices to the clusters during training. Also, the study in~\cite{albaseer2023fair} proposed a client selection and scheduling strategy that utilizes network heterogeneity to schedule clients based on their round latency to accelerate the convergence of models, satisfy fairness, and exploit the reuse of bandwidth.

Despite significant research on the CFL approach, the studies above primarily concentrated on applying CFL in mobile edge networks, so bridging a research gap in exploring CFL in HWNs as a more realistic network architecture is imperative. In addition, critical challenges of CFL in HWNs extend beyond system and data heterogeneities to include resource consumption issues, which arise from transmitting local models to the edge servers-connected cloud in every round. As a result, a hybrid framework incorporating client selection and model aggregation is essential for establishing an effective CFL system within HWNs.
 \begin{figure}[t]
\centering
  \includegraphics[width=1\linewidth]{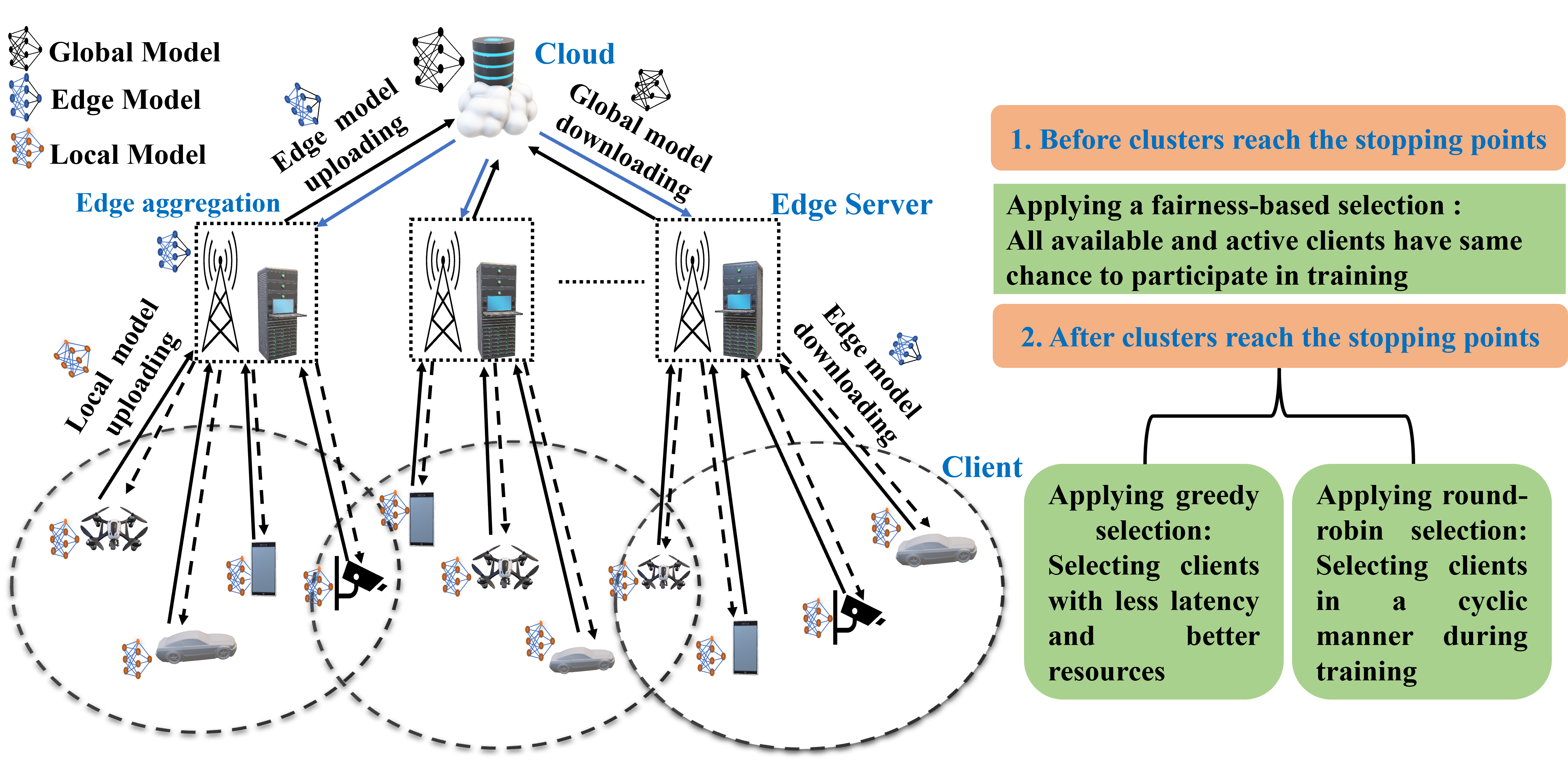}
\caption{{The system model  where the CFL is implemented at HWNs.}}
\label{fig:figure1}
\end{figure}
\begin{table}[h]
\centering
\footnotesize
\caption{Main Symbols }
\label{Tab:Notation}
\begin{tabular}{|p{1cm}|p{6cm}|} 
 \hline

$N$	& number of clients in the HWN \\ \hline
$n$	& index for client $n$ where $n \in N$\\\hline
$K$	& number of edge servers in HWN \\\hline
$k$	& index for edge  server $k$ where $k \in K$\\\hline
$\mathcal{D}_n$ & the local data for $n$-\textit{th} client\\\hline
${\mathcal{D}}$ & the total data samples for all client in the HWN\\\hline
$\omega^n_r$ & the parameters of local  model of $n$-\textit{th} client at $r$-\textit{th} round \\\hline
$\omega^k_r$ & the parameter of edge  model of $k$-\textit{th} edge server at $r$-\textit{th} round  \\\hline
$\omega_r$ & the global model parameters of  the cloud at $r$-\textit{th} round \\\hline
$f_i(\omega_r)$ & the loss function for single data sample \\\hline
$F_n(\omega_r)$ & the local loss function at $r$-\textit{th} HFL round \\\hline
$\nabla F_n(\omega_r)$ & the local gradient for $n$-\textit{th} client\\\hline
$\Gamma_n$ & number of local updates for $n$-\textit{th} client \\\hline
$\eta$ & learning rate \\\hline
$L$ & number of epochs \\\hline
$ SC^k_{n,n'}$& the cosine similarity between $n$ and $n'$ clients in the $k$-\textit{th} edge server \\\hline
$b$ & batch size \\\hline
$\mathcal{S}^k_r$ & the active and available selected clients for edge server $k$ at $r$-\textit{th} round\\\hline
$t^{cmp}_{k,n}$ & local computation time of $n$-\textit{th} client\\\hline
$f_n$ & the local computation frequency at $n$-\textit{th} client\\\hline
$B_n$ & number of cycles required  to process one sample \\\hline
$J(n)$ & the distribution of data of client $n$\\\hline
$T_{budget}$ & the total time budget for the entire training process\\\hline
$T_r$ & deadline time for $r$-\textit{th} round via cloud \\\hline
$\beta^k_n W_k $ & the total allocated bandwidth for $n$-\textit{th} client by edge server $k$\\\hline
$t^{com}_n$ & the transmit time to upload the update to the edge server $n$\\\hline
$r_n,r_k$ &  the achievable data rate required by the $n$-{th} client and $k$-{th} edge server, respectively \\\hline
$\mathbf{h}_n$ & channel gain between the $n$-{th} client and base-station associated with edge server $k$\\\hline
$P_n$ & the $n$-\textit{th} client transmit power \\\hline
$z$ & model size \\\hline
$T^{edge}_{\mathcal{S}_k}$ &  time delay under $k$-\textit{th} edge server with subset of clients$\mathcal{S}_k$\\\hline
$b^k_{n,r}$ & binary variable decides whether a client is selected (one) or not (zero). \\\hline
$\epsilon_1,\epsilon_2$ &  predefined thresholds to control clustering process\\
\hline
\end{tabular}
\end{table}

\section{SYSTEM MODEL}
\label{Sys_model}
As illustrated in Fig. \ref{fig:figure1}, we consider a system model that comprises a cloud server,  a set of edge servers,  which form mobile edge networks, denoted as $\mathcal{K}=\{k:k=1,...., K\}$, $K=|\mathcal{K}|$, and set of clients $\mathcal{N}=\{n: n=1,..., N\}$, $N=|\mathcal{N}|$. It is worth mentioning that in the context of vehicular networks and intelligent transportation systems, edge servers are commonly referred to as Roadside Units (RSUs). Each client owns its local dataset, $\mathcal{D}_n$, to train model parameters locally and then send them to its edge server in the mobile edge network. The number of data samples in each $\mathcal{D}_n$ is $D_n= |\mathcal{D}_n|$, with $\mathcal{D}_n$ representing  the following input-output pairing: $(\{{x}^{(n)}_{i,d}\in \mathbb{R}^d,~ y^{(n)}_{i}\}_{i=1}^{D_n})$, where ${x}^{(n)}_{i,d}$ is the $d$-dimensional input data vector at the $n$-\textit{th} client and $y^{(n)}_{i}$ is the corresponding label associated with ${x}^{(n)}_{i,d}$. Each $k$-\textit{th} edge server holds a subset of clients (i.e., $\mathcal{N}_{k}\subseteq \mathcal{N}$), in which the edge servers are physically located in close proximity to $\mathcal{N}$ clients to coordinate the training process. Afterward, each $k$-\textit{th} edge server aggregates the model parameters and averages them to update its edge model. In return, the cloud collects and averages the edge models from all edge servers to update a global model. During every HFL round, both the cloud and edge servers impose a deadline constraint to synchronize the updates, mitigating potential delays. For ease of presentation, Table \ref{Tab:Notation} summarizes the key symbols used throughout this paper.
\subsection{HFL  Model}
 Within the HFL implemented in HWNs, the cloud aims to capture different patterns from various mobile edge networks. This enables collaborative knowledge sharing among these networks through designing a shared global model, $(\omega_r)$. To this end, the cloud initiates a random global model $(\omega_\circ)$ and forwards it to the edge servers-connected clients for local training. After completing their computations, clients upload their model parameters ($\omega^n_r$)  to their edge servers. In this regard, a synchronized process is used to ensure that all clients, including slower-performing ones (i.e., stragglers), complete their training before uploading their local models to the edge servers. After that, the edge servers aggregate and average local models to update their edge models ( $\omega^k_r$). Finally, the cloud aggregates and fuses all the edge models to form a new version of the global model ($\omega_r$). This process is repeated every global round until the global model converges to the optimal solution.   It is worth noting that the participating clients aim to minimize the loss function using a local solver, i.e.,  a stochastic gradient descent (SGD) optimizer. Specifically,  each client can employ a minibatch SGD in which the batches are used to conduct several updates in a single local epoch. Mathematically speaking, the local training process is defined as:
\begin{equation}
    \Gamma_n=L\frac{|\mathcal{D}_n|}{b},
    \label{No.iterations}
\end{equation}
where $L$ is the number of local epochs to train the local model, and $b$ refers to the batch size determining the number of samples used in one local iteration. During the training, every client aims to fit its local model to accurately predict the corresponding output $y_i$ for a given input $x_i$, effectively minimizing the corresponding loss function $f_i(\omega_r)= l(x^{(n)}_i,y^{(n)}_i; \omega_r)$ by minimizing the error between the actual and predicted labels in every $r$-\textit{th} round.
Therefore, the local loss function of $n$-\textit{th} client is obtained as:
\begin{equation}
    F_n(\omega_r):=\frac{1}{D_n} \sum_{i \in \mathcal{D}_n} f_i(\omega_r).
\end{equation}
Ultimately, the HFL process aims to minimize the total loss function over all clients' datasets $\mathcal{D} = \bigcup_{n} \mathcal{D}_n$ throughout all the mobile edge networks as stated:
\begin{equation}
    \underset{\omega}\min F(\omega,D)=\sum^{N}_{n=1} \frac{D_n}{D} F_n(\omega_r,\mathcal{D}_n)
\end{equation}
\subsection{Clustered Federated Learning (CFL)}
The HFL assumption is violated in realistic applications due to non-IID data distributions of the clients and limitations in the model's complexity. To avoid these challenges, CFL extends this assumption by aiming to generalize it to a group of clients with similar data distributions. This allows clients with similar data to collaborate and benefit from one another while minimizing harmful interference among clients with dissimilar data distributions. Accordingly, the following assumption for CFL is introduced in \cite{sattler2020clustered} as follows:\\
\\
\begin{figure}[t]
\centering
  \includegraphics[width=0.85\linewidth]{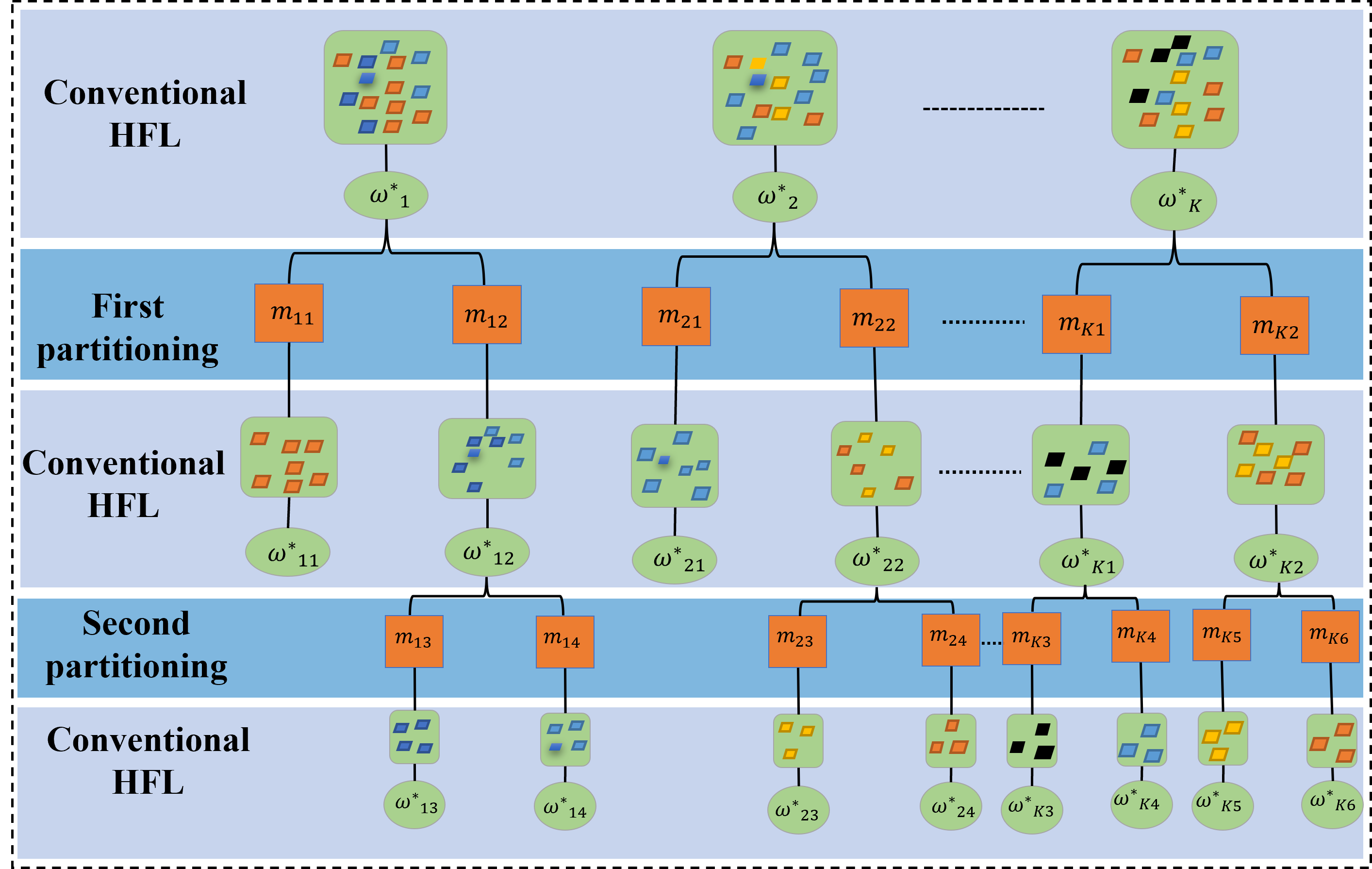}
\caption{Illustration for  CFL in HWN systematically partitions clients into several clusters.}
\label{CFL}
\end{figure}
\textbf{Assumption 1. (``CFL"):}\textit{ For each $k$-th edge server in HWNs, there exists a client population partition $\mathcal{M}_k=\{c_{k1},\dots,c_{km}\},~\bigcup_{j=1}^{M_k} c_{kj}=\{1,\dots,N_k\}$, such that every client' subset $c \in\mathcal{M}_k$ meets the standard FL  assumption}.

 Herein, $\mathcal{M}_k$ is a set of clusters for each $k$-\textit{th} edge server where each holds clients with similar data distributions, and $M_k=|\mathcal{M}_k|$ is the number of clusters. The CFL mechanism in HWNs is illustrated in Fig. \ref{CFL} as a parameterized tree structure, in which the conventional HFL model for every edge network is located at the root node and approaches a stationary point, $\omega^*_k$. After that, each $k$-\textit{th} edge server splits the assigned clients into two clusters based on the similarity in their data distributions. Each cluster iteratively reaches the stationary point, such that ($\omega^*_{k1} \forall k \in K$) and ($\omega^*_{k2} \forall k \in K$) are the stationary points for the first and second clusters, respectively. Here, stationary points refer to situations where the HFL model cannot accommodate clients' different interests due to unbalanced and non-IID data settings. In other words, it is a final solution for the HFL objective where no further improvement is observed in the HFL model with the existing data. Upon meeting this point, splitting clients into different clusters is crucial, ensuring that each model captures more nuanced details within the data distribution in its cluster. It is worth noting that while certain specialized models may have reached their stopping points (i.e., all clients have congruent data distributions) and achieved optimal convergence, others require further training and splitting due to the greater diversity in their data distributions. Therefore, clients experience training and splitting iteratively until further splitting is impossible, where all clusters reach their stopping points and models converge to the optimal solution. It is worth mentioning that the cosine similarity, $CS$, between any two clients (i.e., $n$ and $n'$) in the $k$-\textit{th} mobile edge network within HWNs is calculated as follows:

\begin{footnotesize}
\begin{align}
CS^k_{n,n'}&:=CS(\nabla F_n(\omega^*_k), \nabla F_{n'}(\omega^*_k)):=\frac{\langle  \nabla F_n(\omega^*_k),\nabla F_{n'}(\omega^*_k)\rangle}{\parallel\nabla F_n(\omega^*_k) \parallel \parallel\nabla F_{n'}(\omega^*_k)\parallel} 
\nonumber \\ &= 
\begin{cases}
1, & \text{if $J(n)=J(n')$}\\
-1, & \text{ if $J(n)\neq J(n')$},
\end{cases}
    \label{similarityformula}
\end{align} 
\end{footnotesize}
where $J(n)$ and $J(n')$ are the  distributions of local data for $n$ and $n'$ clients, respectively. Accordingly, the following formulas represent the accurate bipartitioning: $c_{k1}=\{n| CS^k_{n,0}=1\}, \quad c_{k2}=\{n| CS^k_{n,0}=-1\}$. To perform splitting in any $k$-\textit{th} mobile edge network, two crucial conditions should be satisfied~\cite{sattler2020clustered}: 1) the solution approaches the stationary point of the HFL objective, which can be expressed as:
\begin{equation}
  0  \leq \Biggl\|\sum_{n\in \mathcal{N}_k}\frac{D_n}{|D|} \nabla_\omega F_n(\omega^*_k)\Biggr\| < \varepsilon_1,
  \label{cond1}
\end{equation}
and 2) the individual clients are far  from the stationary point of their local loss if  the following condition is satisfied:
\begin{equation}
\max_{n \in \mathcal{N}_k} \left\| \nabla_\omega F_n(\omega^*_k)\right\|>\varepsilon_2>0,
\label{cond2}
\end{equation}
where $\varepsilon_1$ and $\varepsilon_2$ are predetermined hyper-parameters for controlling the clustering task (i.e., determining when to perform splitting for clients). In our proposed approach, we use $\varepsilon_1$ and $\varepsilon_2$ to fine-tune client clustering based on application needs and data heterogeneity. For example, $\varepsilon_1$ is used in the CFL approach to evaluate the clustering quality as the learning process approaches the stationary point, while $\varepsilon_2$ is determined based on client availability and data heterogeneity, controlling how the CFL groups clients into clusters during the clustering mechanism. Precisely, setting $\varepsilon_1$ too large could lead to early clustering, mistakenly combining diverse data distributions, thus reducing model performance. Similarly, setting $\varepsilon_2$ too high can create more generalized clusters, compromising the model's ability to capture and learn from the detailed variations within the data. \\
\textbf{Observation~1:} \textit{If the data distribution for clients is congruent, the partition will not be executed.  This is because when all clients share the same data distribution, the above conditions cannot be met, thereby turning back to the standard FL with one model.}
\subsection{Models for Localized Computation and Communication}
The proposed approach is evaluated from the computational and communicational aspects across the three layers of HWNs, namely, the cloud, edge servers, and clients. The edge server layer performs the evaluation process in three distinct phases.

In the \textbf{first phase}, the computation time and energy for participating clients are analyzed. Let us consider $D_n B_n$ to represent the total number of CPU cycles for the $n$-\textit{th} client to train on its dataset, $\mathcal{D}_n$, during one iteration, where $B_n$ refers to the number of CPU cycles for the client to process a single data sample. Consequently, the computation time for the $n$-\textit{th} client's local iteration is given by $\frac{B_n D_n}{f_n}$, with $f_n$ denoting the CPU speed for the $n$-\textit{th} client. Hence, the computation time for the $n$-\textit{th}  client across $L$ epochs is formulated as follows:
\begin{equation}
    t^{cmp}_{n,k} =L \frac{D_n B_n}{f_n}.
    \label{cmp_time_n}
\end{equation}
Accordingly, the computation energy incurred by the $n$-\textit{th} client to perform $L$ epochs  can be given as:
\begin{equation}
   e_{n,k}^{cmp} =\frac{\alpha_n}{2} f_n^3 t^{cmp}_{n,k},
   \label{energy_cmp_n}
\end{equation}
where $\frac{\alpha_n}{2}$ is the capacitance coefficient related to the energy efficiency of the $n$-\textit{th} client's processor. By inserting (\ref{cmp_time_n}) into the right side of  (\ref{energy_cmp_n}), we obtain:
\begin{equation}
   e_{n,k}^{cmp}=\frac{\alpha_n}{2} (L f_n^2  {D_n B_n}).
   \label{energy_cmp_n_1}
\end{equation}

The \textbf{second phase} involves the communication model for clients. Each mobile edge network within HWNs has a limited number of sub-channels allocated for only a subset of clients (i.e., $S^k_r \subseteq \mathcal{N}_k$), selected every $r$-\textit{th} round.
We utilize orthogonal frequency-division multiple access (OFDMA) techniques to divide the channel into sub-channels. Let us define $W$ as the total system bandwidth divided among all edge servers, with each $k$-\textit{th} edge server receiving a portion of this bandwidth, i.e., $W_k$. Within each $k$-\textit{th} edge server, the  $n$-\textit{th} client is allocated a bandwidth of $\beta^r_{n,k} W_k$.  Here, $\beta^r_{n,k}$ represents the ratio of the bandwidth allocated to the $n$-th client out of $W_k$. Specifically, if  the total bandwidth ,$W$, is split into $N$ sub-channels according to the size of the model $z$, we can obtain the total number of sub-channels as:
\begin{equation}
    N=\frac{W}{z}.
\end{equation}
Hence, the achievable transmission rate for $n$-\textit{th} client to upload its local model to the  $k$-\textit{th} mobile edge network is obtained as:
\begin{equation}
{r^r_n = \beta^r_{n,k}~ W_k~\text {log}_2\left(1 + \frac{| \mathbf{h}^r_n|^2 P^r_n}{ N_\circ }\right)},
\label{transmission_rate}
\end{equation} 
where $| \mathbf{h}^r_n|^2$ represents the channel gain between $n$-\textit{th} client and  $k$-\textit{th} edge server, $ N_\circ$ denotes the complex additive white Gaussian noise (AWGN) power, and $P^r_n$ is the transmission power needed to upload the client's model to the edge server. Therefore, the time required for the $n$-\textit{th} client to upload its local model to the $k$-\textit{th} edge server is defined as follows:
\begin{equation}
    t^{com}_{n,k}= \frac{z}{r^r_{n}}.
    \label{comm_time_n}
\end{equation}
Given the communication time, $t^{com}_{n,k}$, and transmission power, $P^r_n$, of the $n$-\textit{th} client,  the communication energy required for this client to upload a model with size $z$ is defined as:
\begin{align}
    e_{n,k}^{com}=t_{n,k}^{com} P^r_n.
    \label{comm_energy_n}
\end{align}
Again, we have to substitute (\ref{comm_time_n}) to the right-side of (\ref{comm_energy_n}) to achieve:
\begin{align}
    e_{n,k}^{com}=\frac{z P^r_n}{r_{n}}.
\end{align}
As a result, the computation and communication time for the $n$-\textit{th} client in the $r$-\textit{th} round can be obtained as:
\begin{equation}
    T^{tot}_{n,k}=t^{cmp}_{n,k}+t^{com}_{n,k}.
\end{equation}
Similarly, the total energy consumed by each $n$-\textit{th} client in the $r$-\textit{th} round due to computation and communication can be given by:
\begin{equation}
    e^{tot}_{n,k}=  e_{n,k}^{cmp}+ e_{n,k}^{com}.
\end{equation}

The \textbf{third phase} relates to the computation and communication delay and energy of edge servers. Notably, the time allotted to the $k$-\textit{th} edge server is equivalent to the time required by the slowest client to complete its assigned task, which is defined as:
\begin{equation}
 T_{S^k_r}^{\mathrm{edge}} =\underset{n\in S^k_r}\max{~T^{tot}_{n,k}}.
\end{equation}
Furthermore, we calculate the total energy consumed by a set of clients, $S^k_r$, in the $k$-\textit{th} edge server during the $r$-\textit{th} round as follows:
\begin{align}
    E_{S^k_r}^{\mathrm{edge}}=\sum^{S^k_r}_{n=1} e^{tot}_{n,k}.
    \label{Total_energy_edge}
\end{align}
Owing to the substantial computational capabilities of the edge servers, we exclude their time and energy costs for broadcasting the edge and specialized models to the clients from our optimization problem.

As for the cloud layer, all the edge servers upload their edge and specialized models to the cloud to update the global model and capture more behaviors from different mobile edge networks. Hence, the time required for uploading edge and specialized models by the $k$-\textit{th} edge server is given by
\begin{equation}
 T^{\mathrm{cloud}}_{k}= \frac{z}{r^r_{k}},
\end{equation}
 where $r^r_k$ is the achievable transmission rate for the $k$-\textit{th} edge server to upload its edge and specialized models. Note that the cloud aggregates these models and fuses them to create a new version of the global model $\omega_r$. Accordingly, the energy consumption incurred by the $k$-\textit{th} edge server  to upload edge  and specialized models to the cloud  can be derived as:
\begin{align}
    E_{k}^{\mathrm{cloud}}=T^{\mathrm{cloud}}_{k} P_k=\frac{z P_k}{r_{k}}.
\end{align}
Finally, the total system's delay and energy in the $r$-\textit{th} global round can be written as follows:
\begin{equation}
    T_r=\underset{k\in \mathcal{K}} \max (T^{\mathrm{cloud}}_{k}+T_{S^k_r}^{\mathrm{edge}}),
\end{equation}
\begin{align}
    E_r=\sum_{k=1}^{K} (E_{k}^{\mathrm{cloud}}+E_{S^k_r}^{\mathrm{edge}}).
\end{align}
\section{Problem formulation}
\label{problem_form}
While the superior performance of CFL in the mobile edge networks can be achieved by including all participants in the training every round, such an approach is impractical due to two-level model aggregations as well as resource limitations in HWNs. On the other hand, using random selection will fail to capture all the incongruent data distributions among clients. Therefore, these challenges bring about the need to maintain HWN constraints while achieving the desired performance, specifically, efficient specialized models for all clusters in a hierarchical manner. Thus, ensuring the accurate splitting of clients with heterogeneous data distribution and resources is crucial in mobile edge networks. For this purpose, we seek to adjust the CFL setting by performing splitting at the initial stages of training, considering all network constraints. We aim to accelerate the convergence rate and reduce resource consumption. Thus, we need to design an ideal framework that encompasses both the two-level model aggregations and the two-phase client selection. This would enable the creation of tailored, specialized models across different mobile edge networks with minimal associated costs.

As stated, we aim to obtain the optimal model aggregation scheme with two-phase client selection that achieves the optimal set of $M$ models for all edge servers in HWNs, \{${\omega_m},m=1,.., M\}$. The goal is to achieve this within the total time budget ($T_{budget}$) while minimizing the loss function,$F(\omega_m)$, as follows:
\begin{equation}
\omega_{m}=\operatorname*{arg~min}_{\omega_m \in \{\omega_r^{S_{r}} : r=1,...,R\}} F(\omega_{m}).
\end{equation}
Let $R$ denote the total number of global rounds required for training within the time budget $T_{budget}$, and the selected scheduling sets for all rounds denoted as $S_{r}=[S_1, S_2,..., S_R]$, and the deadline for every round is $T_{[R]}=[T_1, T_2,...T_R]$.  The optimization problem for all clusters among different mobile edge networks can be expressed as follows:

\begin{subequations}
\footnotesize

\label{eq:OptmizedProblem1}\vspace{-.5em}
\begin{align}
\textbf{P0: } \quad \mathop{\min}_{
\omega^*,R_{\mathrm{agg}}, R,T_{[R]},S_{[R]}} \:& \sum_{m=1}^{M} F(\omega_{m})
\tag{\theequation}  \\
\textrm{s.t.} \quad 
& F(\omega_{m}) - F(\omega^*_{m}) \le \varepsilon,~\forall m \in \mathcal{M},
\label{eq:convergence_constraint}\\
& \sum_{r}^R T_r(S_r)\leq T_{budget},~r\in \forall [R],
\label{total_time}\\
&  b^k_{n,r}T^{tot}_{n,k} \leq T_{S_r^k}^{\mathrm{edge}}~\forall r,~\forall k,~\forall n \in S_r^k,
\label{Deadline_one_client}\\
& T_r=\max (T_{k}^{\mathrm{cloud}}+T_{\mathrm{S_k}}^{\mathrm{edge}})~\forall k,~\forall r,
\label{eq:FL_time_constraint}\\
& T_{S^k_r}^{\mathrm{edge}} =\max\{ b_{n,r}{~T^{tot}_{n,k}}\} \quad \forall r,\forall n,
\label{eq:FL_time_constraint_each_edge_server}\\
&  S^k_r\subseteq \mathcal{N}_k~\forall r,~ \forall k \in \mathcal{K},
\label{eq:resource_allocation}\\
&   S^k_r \cap S^{k'}_r =\phi,~\forall r, \forall k, k' \in \mathcal{K},~k \neq k',
\label{eq:No_shared_clients}\\
&  b^k_{n,r} \in \{0,1\}.
\label{indicator}
\end{align}
\label{Optim_prob_P0}
\end{subequations}
Further, $R_{\mathrm{agg}}$ denotes the aggregation round at which edge models are collected in the cloud. In problem \textbf{P0}, constraint (\ref{eq:convergence_constraint}) ensures the convergence of the specialized model for each cluster, enabling fairness among groups by assigning each an optimal set of model parameters. Constraint (\ref{total_time}) is to ensure that time consumed during the CFL  process by all selected clients does not exceed the pre-determined time budget, $T_{budget}$ for all rounds while constraints (\ref{Deadline_one_client}) and (\ref{eq:FL_time_constraint}) respectively, impose the time limit of each client for a given round. In parallel, constraint (\ref{eq:FL_time_constraint_each_edge_server}) specifically stipulates the deadline for each edge server per round. Constraints (\ref{eq:resource_allocation}) and (\ref{eq:No_shared_clients}) guarantee that each edge server only associates with available participants during each round, with each participant exclusively dedicated to one edge server. The binary indicator in constraint (\ref{indicator}) distinguishes between selected $(b^k_{n,r}=1)$ and non-selected $(b^k_{n,r}=0)$ participants for training.

One can observe that problem, $\mathbf{P0}$, is an intractable optimization problem due to the necessity of determining how $R$, $R_{\mathrm{agg}}$ and $S_{[R]}$ influence each cluster model's weight vector (i.e., $F(\omega_m)$). Also, an optimal model, $\omega_m$, necessitates careful client clustering based on congruent data distributions. Hence, an iterative solution for $F(\omega_m)$ concerning $R_{\mathrm{agg}}$, $R$ and, $S_{[R]}$ is proposed in Section \ref{proposed_sol}. Given the two-level model aggregation, the dynamic nature of wireless channels, and varying computational time for each client per round, the optimal client selection and suitable model aggregation present complexity. In fact, we use CFL as a recursive technique to select and schedule clients, aiming to account for incongruent data distributions and create specialized models accordingly. 
\section{Proposed Solution}
\label{proposed_sol}
This section introduces our proposed solution to solve problem $\mathbf{P0}$ through integrating client selection and model aggregation in HWNs. Due to the limited resources in the edge devices as well as the two-level model aggregation for different specialized models, we aim to design a framework that achieves better-fit models among different mobile edge networks and optimizes client selection with each edge network itself. Accordingly, we aggregate the edge and specialized models in the cloud using the round-based and split-based model aggregation schemes. In the round-based model aggregation, the cloud instructs the edge servers to upload their edge and specialized models when the global rounds reach the predetermined aggregation rounds, $R_\mathrm{agg}$. In the split-based model aggregation, the edge servers notify the cloud to gather the models once at least one edge server executes a single split.

For the client selection, we have a two-phase process to ensure that the resulting models capture the whole data distribution among all mobile edge networks. In the first phase of our client selection mechanism, we apply the fairness-based selection in which all active and available clients in each edge network have an equal opportunity to participate in training,  contributing to more accurate clustering. In the second phase, if a given cluster reaches the stopping point as in~(\ref{stopping}),  we apply two other client selection methods based on greedy and round-robin algorithms to ensure optimized resource consumption. Adopting greedy and round-robin algorithms after clusters reach their stopping points is motivated by two key factors. First, at the stopping point, all clients within a cluster have the same data distribution; thus, selecting only one client with a better resource and less latency to represent this cluster is an optimal strategy not only to perform further training and improvement but also to reduce the training time and resource consumption significantly. Second, there is no guarantee that clients with better resources and fewer latencies will continue training at the same performance level after a certain number of rounds (i.e., they may face performance degradation); thus, alternating between clients (via round-robin selection) ensures equitable resource usage, preventing the training burden from falling on a single client. Mathematically, for each $k$-\textit{th} edge network, let $S^k_r$ represent all clients in clusters that have yet to approach the stopping point, and let $S^k_\mathrm{grdy}$ denote a subset of clients from other clusters that have reached the stopping point. These conditions can be formally expressed as follows:
\begin{align}
   S^k_\mathrm{grdy}=&\{j\leftarrow\mathrm{arg}~\underset{j \in m}\min~\{ T_j^{total}~\forall j \in m\\ \nonumber
   &|\forall m| \underaccent{n\in m}\max ||\nabla_{\omega_k} F_n(\omega_k^*)||\leq \epsilon_2\}\}
\label{eq:greedy}
\end{align}
\begin{equation}
    S^k_r=S^k_r \cup  S^k_\mathrm{grdy}.
    \label{full}
\end{equation}
It can be observed from (18) that applying the greedy and round-robin client selection methods only occurs for the clusters reaching the stopping point. It is worth highlighting that the proposed approach not only effectively tackles unbalanced and non-IID data but also inherently addresses system heterogeneity using two-phase client selection and two-level model aggregation schemes. In the second phase of client selection, we use greedy and round-robin algorithms for efficient scheduling based on system heterogeneity, ensuring optimal client participation. In the greedy selection, clients with less latency and better resources are prioritized for continued training, enhancing model convergence, accelerating model updates, and reducing resource consumption. Simultaneously, the round-robin algorithm ensures equitable participation by fairly distributing the training load among all clients, thus mitigating bias towards less capable clients, preventing network congestion, and enhancing bandwidth efficiency. For the two-level model aggregation, we not only minimize network traffic and transmission delays, which is beneficial for clients with limited bandwidth but also accelerate model updates through local aggregation at edge servers. This approach ensures efficient training and swift decision-making, which is crucial for real-time applications, such as IoT devices and intelligent vehicles.

It is worth noting that the proposed approach is tailored for a wide range of applications, particularly emphasizing IoT devices and intelligent vehicle systems, which demand immediate data processing and effective distributed learning. This flexibility makes the proposed approach applicable to real-world applications requiring swift decision-making. By implementing two-phase client selection and two-level model aggregation, our proposed approach has the ability to significantly improve model performance and network efficiency across various HWN applications.
\subsection{Detailed Schemes}
We apply our proposed framework across four different schemes, each innovatively crafted to address the challenges of CFL in HWNs. Detailed descriptions of these schemes are subsequently provided.
 \subsubsection{Fair and Greedy Selection with Round-based Model Aggregation}
 \label{scenario_A}
 As stated before, the $k$-\textit{th} edge server uses a two-phase client selection process; namely,  fairness-based and greedy-based selection algorithms. For model aggregation, the cloud collects all models using the round-based method. The learning process occurs as follows: all the clients are provided with the same opportunity to participate in the training process, in which each client uses its local dataset to train the local model parameters and uploads them to the associated edge server for aggregation. After that, each $k$-\textit{th} edge server carries out two main tasks. Initially, it computes the cosine similarity between clients' gradients to cluster them based on the similarity in their data distributions, providing each cluster with a specialized model that fits well with the data distribution. Subsequently, it averages the local models to update its edge model and returns the edge and specialized models to the clients for further training. This iterative process continues until a particular cluster reaches the stopping point. In this case, the $k$-\textit{th} edge server uses the greedy algorithm to select low-latency, resource-efficient clients for further training. When the designated aggregation rounds, $R_{\mathrm{agg}}$, are met, the cloud collects edge and specialized models from the edge servers, mimicking the process at the edge servers. Precisely, the cloud computes the cosine similarity between specialized models to capture more patterns from mobile edge networks and encourage collaborative learning. Meanwhile, the cloud averages the edge models to update the global model, which then distributes it back, along with the specialized models, to the edge servers and then to the clients for further training. This process iteratively continues until the global and specialized models converge toward the optimal solution.
\subsubsection{ Fair and Greedy Selection with Split-based Model Aggregation}
\label{scenario_B}
The client selection process in this scheme follows the same process of client selection described in scheme \ref{scenario_A}; however, aggregating edge and specialized models in the cloud is achieved via the split-based model aggregation scheme. 
Specifically, all available clients equally participate in local model training, while the edge servers aggregate these local models, compute similarities, update edge models, and return them for further training.
Once the edge servers perform at least one split, the cloud aggregates the edge and specialized models using the split-based model aggregation scheme. Precisely, when the splitting conditions stated in  (\ref{cond1}) and (\ref{cond2}) are satisfied for any $k$-\textit{th} edge server, the $k$-\textit{th} edge server notifies the cloud to aggregate both the edge and specialized models from all mobile edge networks. Again, the cloud repeats the same procedure as in scheme \ref{scenario_A}. When a specific cluster reaches the stopping point, the corresponding edge server applies the greedy algorithm to select the clients with less latency to perform further updates. 
This process is repeated for $R$ rounds until the global and specialized models converge.
\subsubsection{Fair and Round-robin Selection with Round-based Model Aggregation}
\label{scenario_C}
The client selection algorithm differs from the previous two schemes, where the $k$-\textit{th} edge server utilizes round-robin selection for each cluster meeting the stopping point. In contrast, the model aggregation is exactly the same as in scheme \ref{scenario_A}. Specifically, the fairness-based selection is used to select all clients to participate in the early stage of training. Once a given cluster attains the stopping point, the corresponding edge server adopts the round-robin algorithm to select clients for further training. This process ensures a fair distribution of resources and opportunities for all clients in that cluster.  Once the aggregation interval round, $R_{\mathrm{agg}}$, is achieved, the cloud aggregates the edge and specialized models and then follows the same steps as in the previous schemes. This process continues until the global and specialized models converge on the optimal solution, ensuring efficient and effective learning for the  CFL within the HWN.
\subsubsection{ Fair and Round-robin Selection with Split-based Model Aggregation}
In this scheme, selecting the clients follows the same process as in scheme \ref{scenario_C}, and model aggregation follows the same steps as in scheme \ref{scenario_B}. More specifically, Once the split conditions, as specified in (\ref{cond1}) and (\ref{cond2}), are met for any $k$-\textit{th} edge server, this edge server informs the cloud, prompting the aggregation of the edge and specialized models for all mobile edge networks. The cloud performs the same steps as in all previous schemes, and the process continues repeatedly until the global and specialized models converge to the optimal solution.\\

It is worth noting that the round-based model aggregation scheme presents a trade-off between enhancing model performance and managing resource consumption. Assigning a large value to  $R_{\mathrm{agg}}$  enables more efficient use of network resources. However, this requires careful consideration to avoid compromising overall system performance. Alternatively, setting a small value to $R_{\mathrm{agg}}$ enhances model performance and accuracy, albeit leading to an increase in resource consumption. Therefore, selecting $R_{\mathrm{agg}}$ value is critical in maintaining a balanced trade-off between resource consumption and model performance.  Regarding the split-based model aggregation scheme, collecting models in the cloud directly depends on the rate at which clients cluster based on the similarity in their data distributions.  Several factors are crucial in determining this splitting time, including the number of clients involved, the inherent communication latency, and the degree of similarity among the datasets. Hence,  understanding and managing these factors is crucial to optimizing system performance and resource utilization.

Similarly,  we observe a trade-off between the number of selected participants per round and the total number of rounds, a complexity managed by  $T_r$. If initiated with a short $T_r$, the number of participants in $S_r$  decreases, a situation unfavorable for CFL as it could result in slow client partitioning into congruent clusters. This produces unnecessary communication costs owing to the increased number of training rounds necessary for convergence. Consequently, we propose initializing the training with a long $T_r$. As accurate clustering is achieved,  $T_r$ is reduced either by cyclically selecting the clients  (round-robin selection) or selecting the most efficient clients with less latency (greedy selection) from clusters that reach the stopping point, which  can be delineated as follows:
\begin{equation}
\max_{n \in c} \left\| \nabla_\omega F_n(\omega^*_k)\right\|<\varepsilon_2.
\label{stopping}
\end{equation}
\begin{algorithm}[t!]
\footnotesize
\caption{CFL with Round-based  Aggregation}
\label{CFL0}
 \KwIn{Number of clients ($N$), initial parameters ($\omega_\circ$), controlling parameters $\varepsilon_1$, $\varepsilon_2 >0$, epochs' number ($L$), and batch size $b$.}
 \KwOut{$M$ specialized models, one  FL global model}
  \textbf{Initialize:}  Set group of  clients $\mathcal{N}=\{\{1,2,\dots,N_1\},\dots,\{1,2,\dots,N_K\}\}$ as initial clustering, initial models $\omega_n \leftarrow 0~\forall n$, and $r=1$ \par
\For{$r=1$ to $R$}
 {\textbf{Edge aggregation} \par
 \For{$e=1$ to $K$}
 {\textbf{//Pre-processing of $k$-\textit{th} edge server}\par
 { $k$-\textit{th} edge server select all active and available clients}\par
$k$-\textit{th} edge server sends $\omega_{r-1}$ to all clients \par
\If{$M>1$}{Edge server picks \par$\{m \in \mathcal{M}|\underset{n\in m}{\max}\parallel \nabla_\omega F_n(\omega^*_k)\parallel<\varepsilon_2>0\}$}
\Else{Schedule the participation schedule for all individuals interested in participating in training}\par
  \textbf{//Participating clients task in parallel}\par
 \For{$n=1$ to $|\mathcal{S}^k_{r}|$}
 {  receive $\omega_{r-1}$\par
 Execute local training $\omega^r_n=\omega^{r-1}_n-\eta\sum_{t=1}^{\tau}\nabla F_n(\omega_n{(t)})$}
 Server aggregates all local models\par
  \textbf{//Post-processing steps performed edge server $k$}\par
$\mathcal{M}_{\mathrm{tmp}} \gets \mathcal{M}$\par
 Edge server finds $F(\omega^r)=\sum^{N}_{n=1} \frac{D_n}{D}F_n(\omega^r)$ and $\omega^r=\sum^{N}_{n=1} \frac{D_n}{D}\omega^r$\par
 \For{$c \in \mathcal{M}$}
 {
      Edge server gets $\triangle\omega_m \gets \frac{1}{|c|} \sum_{n\in c} \triangle \omega_n$\par
       \If{$||\triangle \omega_c||<\varepsilon_1$, $\mathrm{and}$ $\max_{n \in c }||\triangle \omega_n||> \varepsilon_2$}
      {
           $SC^k_{n,n'} \gets \frac{\langle \triangle\omega_n, \triangle\omega_{n'}\rangle}{||\triangle\omega_n||||\triangle\omega_{n'}||}$ \par
           $c_1,c_2\gets\arg\underset{\rm c_1\cup c_2=c}{\rm min} \Bigl(\
\underset{n\in c_1, n'\in c_2}{\max} CS^k_{n,n'}\Bigr)$ \par
           $\gamma_n:=\frac{||\nabla F_{I(n)}(\omega^*)-\nabla F_n(\omega^*)||}{||\nabla F_{I(n)}(\omega^*)||}$\par
     \If{$\max(\gamma_n)<\sqrt{\frac{1-\mathrm{CS}^{\max}_{\mathrm{cross}}}{2}}$}{ $\mathcal{M}_{\mathrm{tmp}} \gets (\mathcal{M}_{\mathrm{tmp}}\setminus c) \cup \{c_1, c_2\} $} }   
 }
 $\mathcal{M} \gets \mathcal{M}_{\mathrm{tmp}}$
 }\par
 \textbf{Edge Servers} return $K$ edge models and $M$ specialized models \par
 \textbf{Cloud aggregation}\par
 {\If{$r$~\%~$R_{\mathrm{agg}}=0$}
 {\If{$\textbf{M}>2$}{Cloud repeats steps in lines  18-28 to compute similarities between $M$ specialized models from different mobile edge networks\newline
 Cloud aggregates edge  and specialized models and updates the global model ($\omega_r$)}
}}$r=r+1$
}
{Cloud returns one global model and $M$ specialized models}
\end{algorithm}
\begin{algorithm}[t!]
\footnotesize
\caption{CFL  with Split-based Aggregation}
\label{CFL1}
 \KwIn{ Number of clients ($N$), initial parameters ($\omega_\circ$), controlling parameters $\varepsilon_1$, $\varepsilon_2 >0$, epochs' number ($L$), and batch size $b$.}
 \KwOut{$M$ specialized models, one  FL global model}
  \textbf{Initialize}  Set group of  clients $\mathcal{N}=\{\{1,2,\dots,N_1\},\dots,\{1,2,\dots,N_K\}\}$ as initial clustering, initial model $\omega_n \leftarrow 0~\forall n$, and $r=1$\par
  \textbf{Edge aggregation} \par
 \For{$r=1$ to $R$}
 { \If{$M>1$}{Edge server picks\par$\{m \in \mathcal{M}|\underset{n\in m}{\max}\left\| \nabla_\omega F_n(\omega^*_k)\right\|<\varepsilon_2>0\}$}
\Else{Arranging the participation schedule for all individuals interested in participating in training}\par
\textbf{If edge servers perform at least one split}\par
  \textbf{Cloud aggregation} is occurred\par
\If{$\textbf{M}>2$}{Cloud repeats steps in lines  18-28 in Algorithm \ref{CFL0} to compute similarities between $M$ specialized models from various mobile edge networks\par
Cloud aggregates all models and updates the global model ($\omega_r$).}
$r=r+1$}
Cloud returns one global model and $M$ specialized models
\end{algorithm}
If condition (\ref{stopping}) is met, all clients within cluster $c$ have congruent data distribution, and the CFL  process terminates for that cluster, returning the optimal solution $\omega^*_k$. Otherwise, the data distribution for clients is incongruent, and the $k$-\textit{th}  edge server splits the clients utilizing  (\ref{similarityformula}), (\ref{cond1}), and (\ref{cond2}). According to the splitting criteria of CFL, there are two types of cosine similarities in our proposed approach. The first type is the cross-cosine similarity, which is used to measure the similarity between two clients from different clusters and can be written as:
\begin{equation}
CS^{\max}_{\mathrm{cross}}=\max_{n \in c_1,n' \in c_2}CS(\nabla_\omega F_n(\omega^*), \nabla_\omega F_{n'}(\omega^*)).    
\label{corss_sim}
\end{equation}
 The second type is the intra-cosine similarity which measures the similarity between clients within the same cluster, which can be expressed as:
\begin{equation}
CS^{\min}_{\mathrm{intra}}={\min_{n, n'\atop J(n)=J(n')}CS(\nabla_\omega F_n(\omega^*), \nabla_\omega F_{n'}(\omega^*))}.    
\label{intra_sim}
\end{equation}
As a result, CFL aims to create clusters that maximize intra-cluster similarities and minimize cross-cluster similarities. In the worst case of CFL, it is always correct to partition clients into two clusters if the separation gap between the two types of similarities is greater than zero. Otherwise, there is no need for splitting. Thus, the separation gap can be defined as:
\begin{equation}
    G(SC)=SC^{\min}_{\mathrm{intra}}-SC^{\max}_{\mathrm{cross}}.
\end{equation}
Algorithms (\ref{CFL0}) and (\ref{CFL1}) present the detailed steps of the proposed framework, which comprise the scheduling technique, model aggregation, and the CFL mechanisms in the HWNs. The input parameters include $N$, $\varepsilon_1, \varepsilon_2>0$, $L$, and $b$. For initializing these algorithms, all clients are clustered before training into $K$ groups according to the cost (distance-based association). As in Algorithm (\ref{CFL0}), lines (8-9) determine whether the clustering process by the CFL has been initiated across all edge servers based on the similarity check. Once clustering occurs, the $k$-\textit{th} edge server evaluates each cluster to identify if a stopping point has been reached. If so, the edge server proceeds with client selection for training, employing either the greedy or round-robin algorithms. Selecting the algorithm relies on the intended goal: model performance efficiency guides the use of the greedy algorithm, whereas fairness in resource distribution directs applying the round-robin approach. Conversely, for those clusters that do not reach the stopping point, the $k$-\textit{th} edge server must select all active and available clients to complete the training. At the stopping point, it is essential to note that all clients within the cluster share a congruent data distribution.  Therefore, based on the specific scheme, either the greedy or round-robin selection method can be executed for subsequent updates (lines 7–12).
Line (14) indicates that the $k$-\textit{th} edge server sends the most recent models to the selected clients.  The selected clients carry out standard FL updates in lines (13-15), returning these updates to the edge server that aggregates (line 15) all updates. Suppose the last cluster's model approaches the stationary points (lines 18-27). In that case, the $k$-\textit{th} edge server executes the grouping technique to cluster the clients based on their local data distribution similarity. Otherwise, the $k$-\textit{th} edge server will continue to use the conventional FL algorithm. This iterative process continues until each model approaches its stationary point, catching all incongruent data distributions. After aggregating all edge and specialized models (line 34 in Algorithm(\ref{CFL0}) and line 12 in Algorithm (\ref{CFL1})), they are forwarded to the cloud for further processing. Collecting these models in the cloud occurs either based on round-based (Algorithm \ref{CFL0}) or split-based (Algorithm \ref{CFL1}) schemes. The cloud in both algorithms repeats the same procedure as on the edge servers by computing the similarities among the received models from different mobile edge networks to enable collaborative training and generalize the models to fit new data.
\subsection{Algorithm Complexity Analysis}
Here, we introduce an analysis of the proposed algorithms,  focusing on their computational and communication complexities, networking aspects, and overall energy consumption.

For Algorithm \ref{CFL0}, we use the round-based model aggregation at the cloud-level of aggregation, while every-round model aggregation is used for the edge-level aggregation. In the edge-level aggregation, the complexity is proportional to the number of participating clients and the size of the model parameters; thus, total complexity across all $R$ rounds and all $K$ edge networks is given as $\mathcal{O}(Rz\sum_{k=1}^{K}S^k)$, where $S_k$ is the active and available clients within the $k$-\textit{th} edge network. In cloud-level aggregation, the complexity is $\mathcal{O}(\frac{R}{R_{\mathrm{agg}}}zK)$. Therefore, the total complexity for both levels of model aggregation over the entire training process can be represented as $\mathcal{O}(Rz\sum_{k=1}^{K}S_k)+\mathcal{O}(\frac{R}{R_{\mathrm{agg}}}zK)$.
The algorithm also integrates two client selection phases: a fairness-based selection to ensure equitable participation early in training and a second phase utilizing greedy and round-robin algorithms for optimized resource use. The complexity of the initial selection phase across all networks is linear, $\mathcal{O}(\sum_{k=1}^{K}S_k)$. In the second phase, employing whether greedy or round-robin, the complexity is notably lower because only a single client is selected for training in each round. Thus, the complexity per cluster is represented as $\mathcal{O}(S^c_k)$, assuming constant time for evaluating each client. Here, $S^c_k$ is the total number of clients within the $c$-\textit{th} cluster that reaches the stopping point. Therefore, the overall complexity for two-phase client selection is given as $\mathcal{O}(\sum_{k=1}^{K}S_k+S^c_k)$.

For Algorithm \ref{CFL1}, the complexity is similar to that in Algorithm \ref{CFL0}, with the only difference being the way of aggregating models at the cloud-level aggregation. Here, we adopt a split-based model aggregation that substantially decreases the complexity associated with uploading models. This reduction is achieved by only initiating model uploads when a split occurs at any edge server. Thus, the complexity of this scheme adjusts to $\mathcal{O}(S_{\mathrm{st}}zK)$, where $S_{\mathrm{st}}$ is the total number of splits across all edge servers during the entire training process.  It is worth noting that incorporating these phases of client selection and levels of model aggregation introduces manageable complexity, which is justified by our experiments' significant improvements in learning performance and resource consumption (i.e., more elaboration is introduced in Section \ref{results}).

\section{Performance Evaluation}
\label{results}
\subsection{Experimental Setup}
This section presents the conducted simulations to evaluate the proposed approach compared to the baselines in terms of learning performance and resource consumption.\\
\textbf{Experimental Parameters:} We use a system's bandwidth of 10 MHz, with 1 MHz for each sub-channel. The channel gain ($\mathbf{h}^r_n$) for each client is modeled with a path loss ($\alpha=g_\circ(\frac{d_\circ}{d})^4$), where $g_\circ=-35$ dB and the distance ($d_\circ=2$ m). Also, AWGN power is set to ($N_\circ=10^{-8}$ watts). The client is $20$~m to $100$~m away from a given edge server. The computation frequency $f_n$ ranges between $1$ GHz and $9$ GHz, and the number of cycles per sample is identical for all clients, i.e., $B_n=20$ cycle/sample. The transmit power $P^r_n$ varies between the upper bound $p^{max}_n=20$ dBm and lower bound $p^{min}_n=-10$ dBm.\\
\textbf{Dataset Preprocessing:}
To measure the effectiveness of our proposed approach, we carry out a series of simulations using the FEMNIST and CIFAR-10 datasets~\cite{gowal2021improving,caldas2018leaf}. These datasets are widely used for handwriting classification and object recognition tasks, respectively, in the FL setting.  Specifically, the CIFAR-10 dataset can be used in autonomous vehicles, where image recognition and interpretation are crucial for safe and efficient operation. FEMNIST explicitly helps classify letters and digits (A-Z, a-z, and 0-9), comprising 244,154 images for training and 61,500 for testing, while CIFAR-10 comprises 60,000 32x32 colored images. Also, we apply these datasets in a non-IID fashion, partitioning them into $\mathcal{J}$ fragments and restricting each client to hold only two random classes. Regarding the model architecture, a convolutional neural network (CNN) classifier is employed for FEMNIST, which is integrated with two hidden layers, while Alexnet, a deep neural network (DNN), is utilized for CIFAR-10. The Rectified Linear Unit (ReLU) activation function is applied to the hidden layers in both learning tasks, while the output layer utilizes the Softmax activation function. For local training, we partition $80\%$ for training and $20\%$ for testing. The simulation parameters are listed in Table~\ref{tab:setuppar} for ease of presentation.
\begin{table}[h]
\footnotesize
	\caption{ List of Simulation parameters}
	\label{parameter}
	\centering
	\begin{tabular}{|c|p{4.5cm}|}%
		\hline
		\textbf{Parameter}&\textbf{Value}\\
		\hline
		No. of clients (N) & 200\\		\hline
        No. training rounds ($R$) & $200$\\		\hline

		 No. of edge servers (K) & 3\\		\hline

		 No. of epochs (L) & 10\\		\hline

		 batch size (b) & 32\\		\hline

		 Learning rate ($\eta$) & 0.01\\		\hline

		System bandwidth (W) & 10 MHz\\		\hline

		 Transmission Power ($p_n$) & [$-10, 20$] dBm  \\		\hline

        AWGN power ($N_\circ$) & $10^{-8}$ W \\		\hline

		 CPU Frequency ($f_n$) & [$1, 9$] GHz\\		\hline

        Cycles Per Sample $(B_n)$ & 20 cycle/sample \\		\hline


		\hline
	\end{tabular}
\label{tab:setuppar}
\end{table}
\\
 \textbf{Performance Metrics and Benchmarks:} Our proposed schemes are critically evaluated against the CFL algorithms, serving as the baselines~\cite{sattler2020clustered,duan2020fedgroup}. The baseline strategy randomly selects clients without accounting for factors such as wireless channel quality, latency in training and transmission, or resource availability. Additionally, edge models and specialized modes are aggregated to the cloud from the edge servers every round. The performance metrics used for this evaluation are the testing accuracy and energy consumption, as illustrated in Figs. \ref{acc_norm}$-$\ref{Energy_split}.
  \begin{figure}[t]
\centering   
     \includegraphics[width=0.7\linewidth]{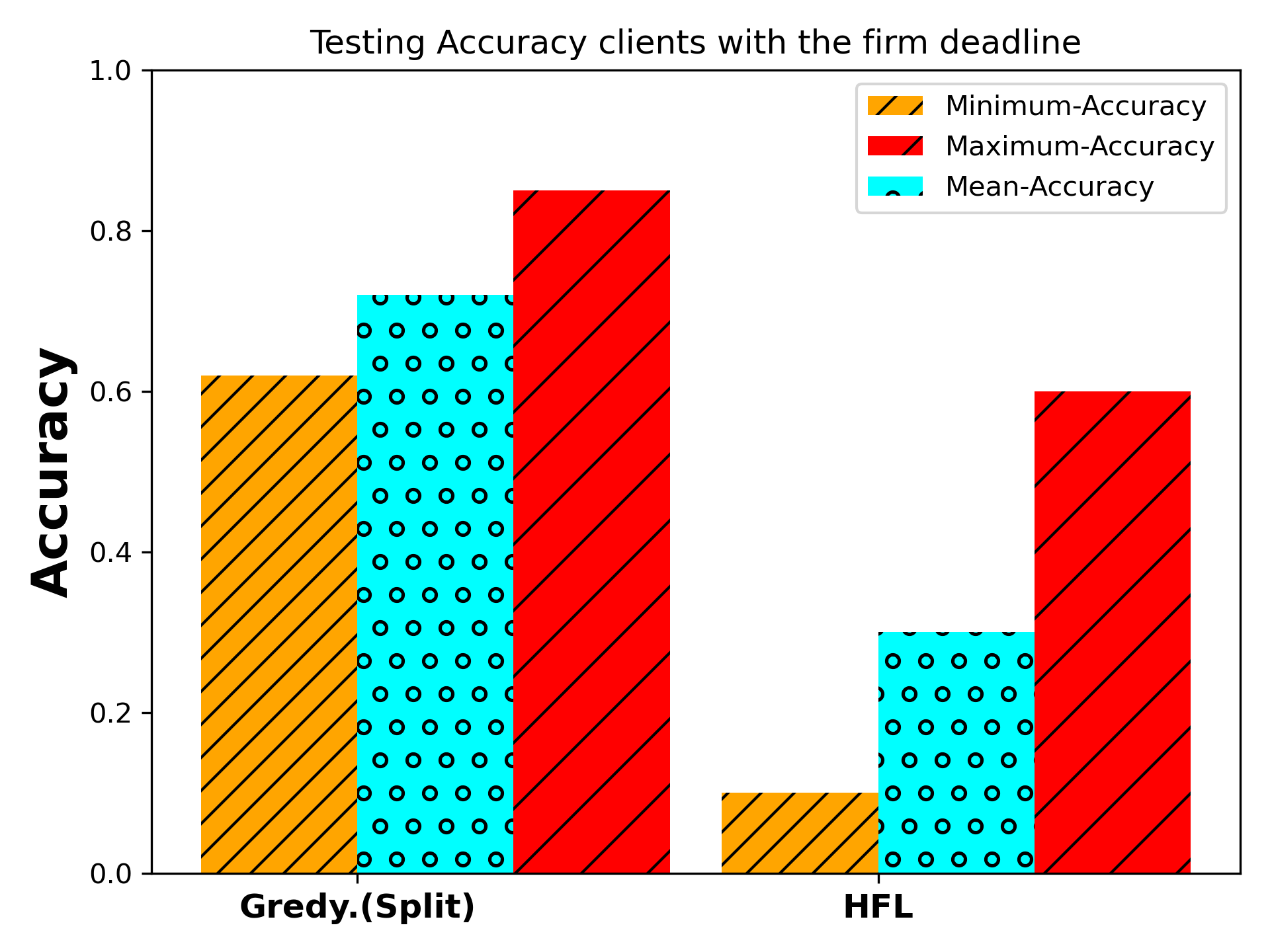}
    \caption{Testing accuracy for fair and greedy selection with split-based model aggregation and HFL (FEMNIST).  }
    \label{HFL with CFL}
\end{figure}
 \begin{figure*}[t]
    \centering  
\begin{subfigure}[b]{0.45\textwidth}
\includegraphics[width=0.9\linewidth]{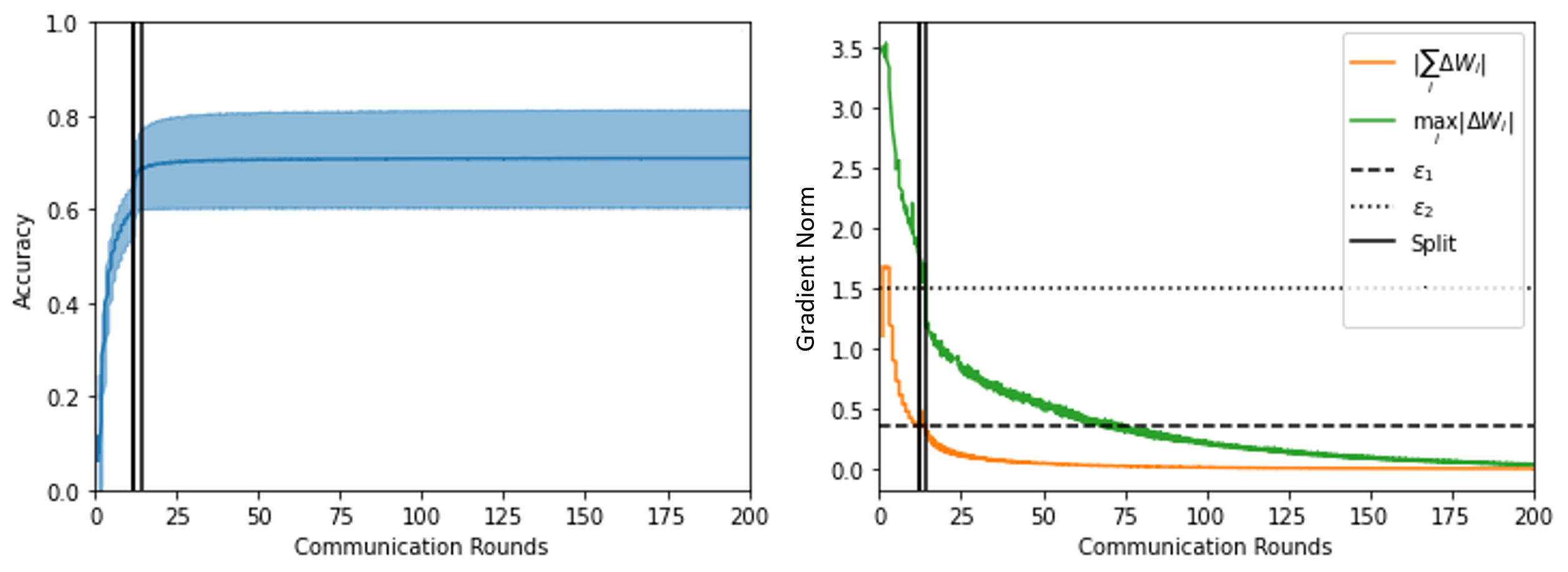}
\caption{Fair and greedy selection with round-based model aggregation algorithm for $R_\mathrm{agg}=5$.}
\label{greedy_r=5}
\end{subfigure}
\begin{subfigure}[b]{0.45\textwidth}
\includegraphics[width=0.9\linewidth]{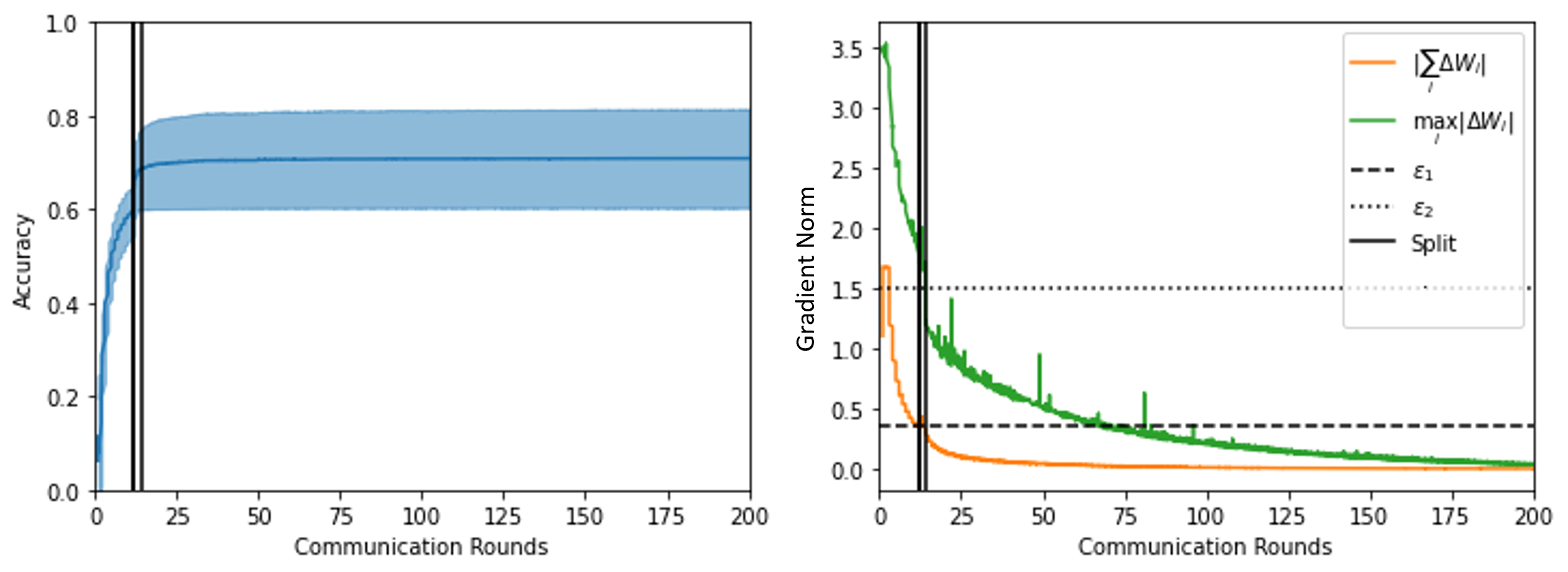}
\caption{Fair and  greedy selection with round-based model aggregation algorithm for $R_\mathrm{agg}=10$. }
\label{greedy_r=10}
\end{subfigure}
\begin{subfigure}[b]{0.45\textwidth}
\includegraphics[width=0.9\linewidth]{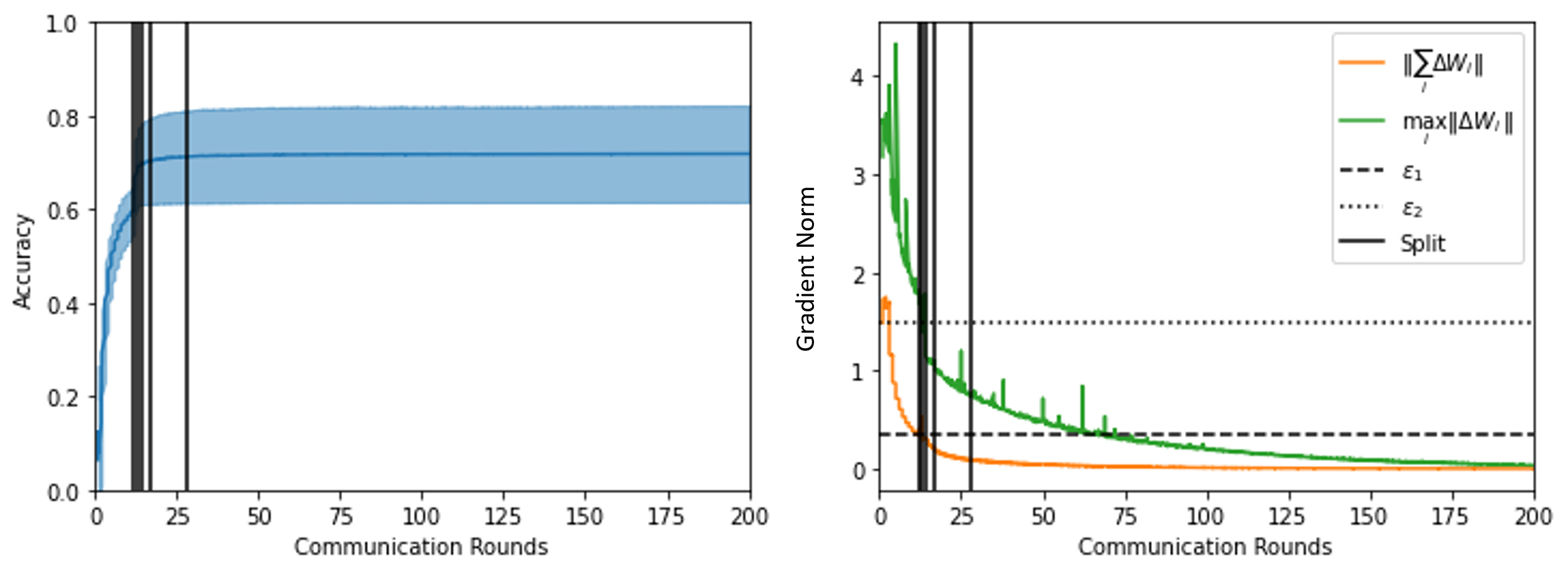}
    \caption{Fair and round-robin selection with round-based model aggregation algorithm for $R_\mathrm{agg}=5$. }
    \label{robin_r=5}
\end{subfigure}
\begin{subfigure}[b]{0.45\textwidth}
\includegraphics[width=0.9\linewidth]{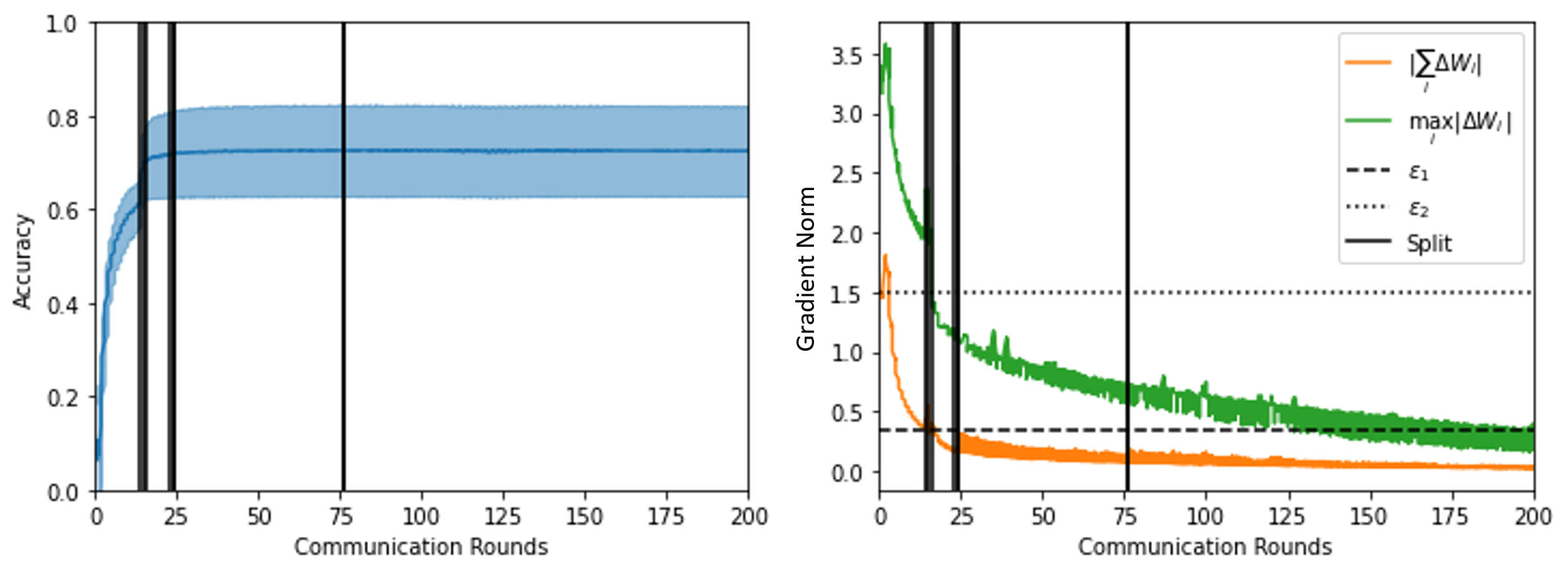}
    \caption{Fair and round-robin selection with round-based model aggregation algorithm for $R_\mathrm{agg}=10$. }
    \label{robin_r=10}
\end{subfigure}
\begin{subfigure}[b]{0.45\textwidth}
\includegraphics[width=0.9\linewidth]{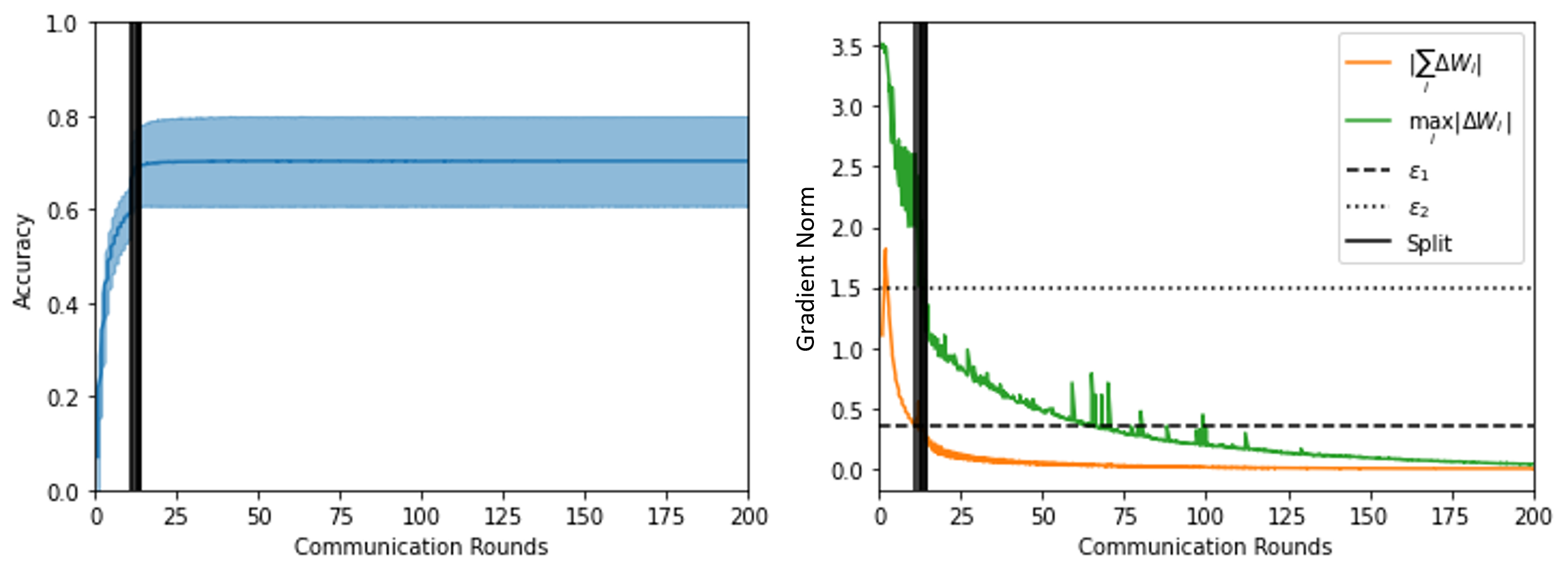}
    \caption{Fair and greedy selection with split-based model aggregation algorithm. }
    \label{greedy_split}
\end{subfigure}
\begin{subfigure}[b]{0.45\textwidth}
\includegraphics[width=0.9\linewidth]{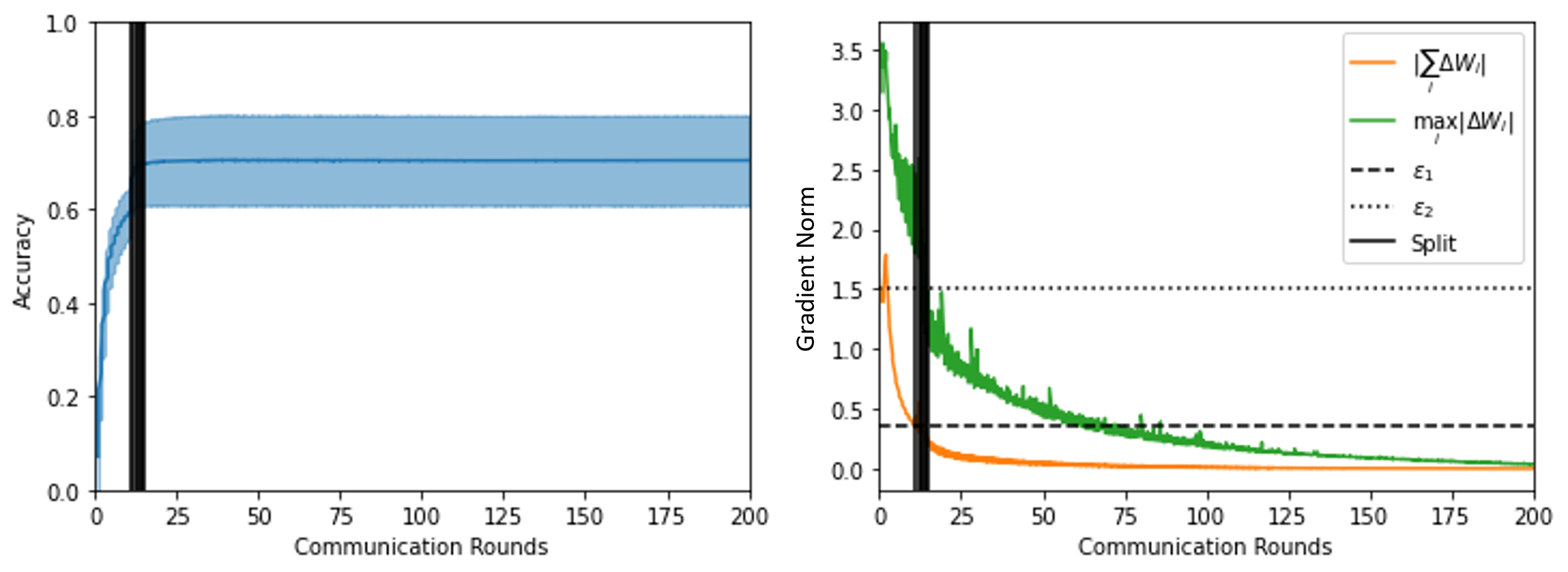}
    \caption{Fair and round-robin selection with split-based model aggregation algorithm.}
    \label{robin_split}
\end{subfigure}
\begin{subfigure}[b]{0.45\textwidth}
\includegraphics[width=0.9\linewidth]{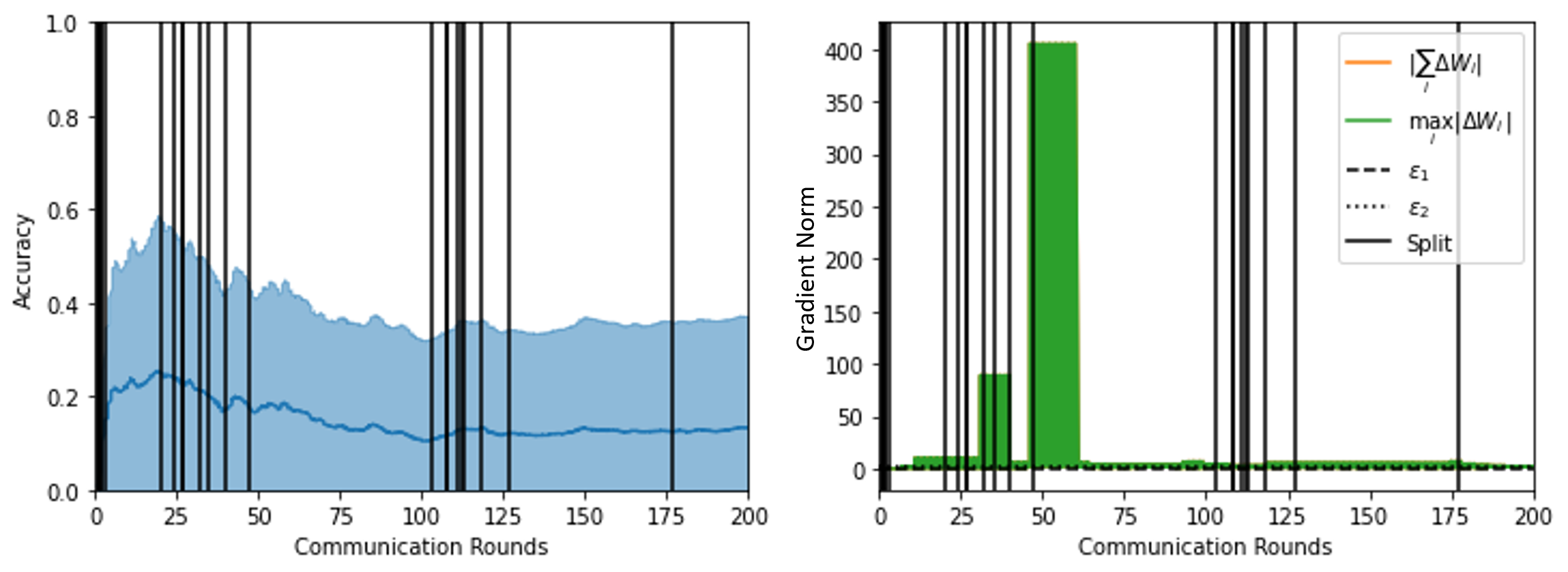}
    \caption{The Baseline.}
    \label{tradi_CFL}
\end{subfigure}
\caption{Average testing accuracy for models of clusters during training rounds for both proposed approach and baselines using FEMNIST dataset}
\label{acc_norm}
\end{figure*}
\begin{figure}[t]
\centering   
     \includegraphics[width=1\linewidth]{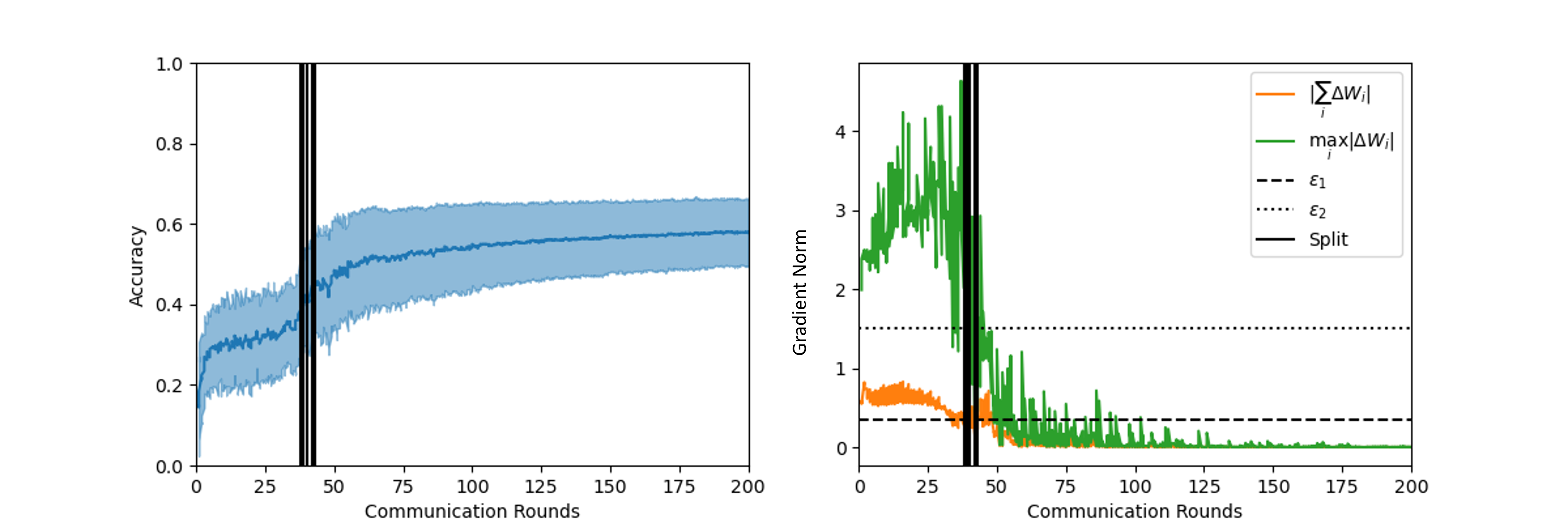}
    \caption{Fair and greedy selection with split-based model aggregation using CIFAR-10 dataset. }
    \label{acc_CIFAR10}
\end{figure}
\begin{figure}[t]
\centering 
\includegraphics[width=1\linewidth]{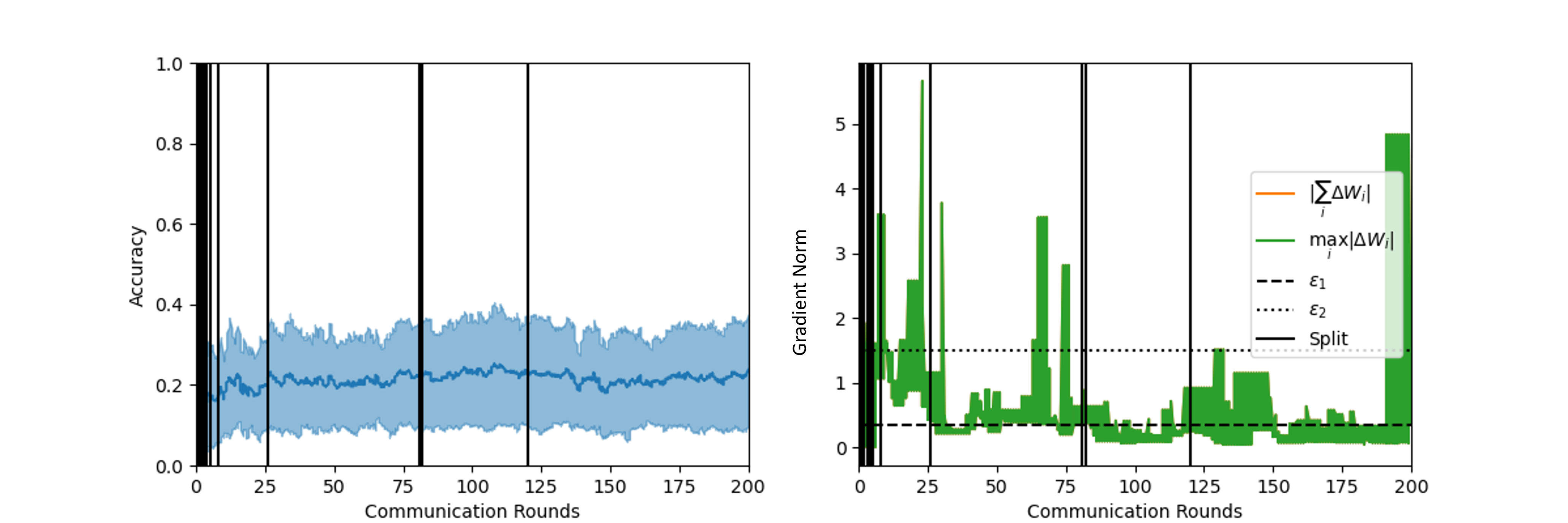}
    \caption{The CFL with random selection (baseline) using CIFAR-10 dataset. }
    \label{CFL_CIFAR-10}
\end{figure}
\begin{figure*}[t]
\centering   
\begin{subfigure}[b]{0.23\textwidth}
\includegraphics[width=0.9\linewidth]{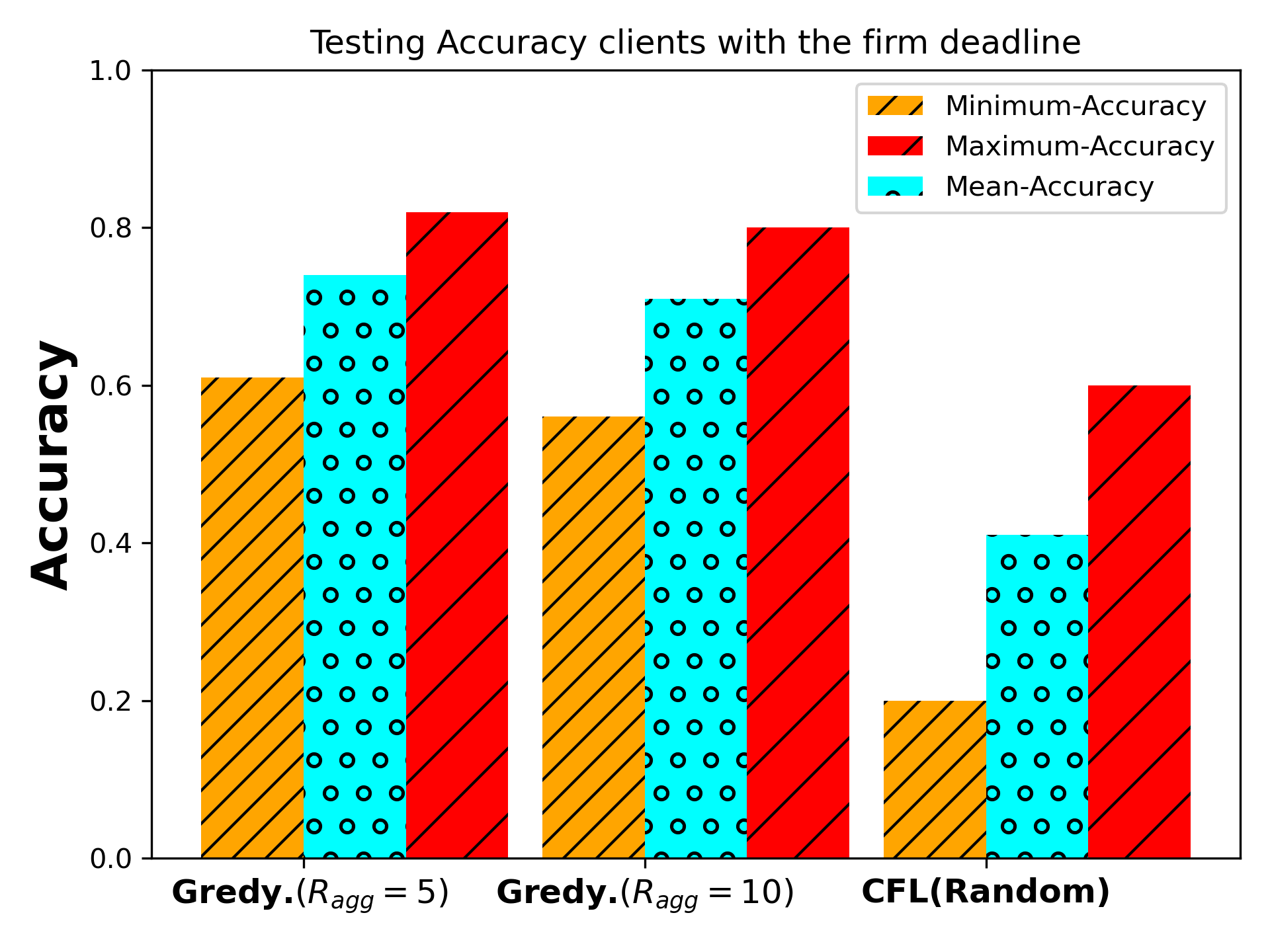}
\caption{ Fair and greedy selection with round-based model aggregation ($R_\mathrm{agg}=5$).}

\label{acc_greedy_r=5}
\end{subfigure}
\begin{subfigure}[b]{0.23\textwidth}
\includegraphics[width=0.9\linewidth]{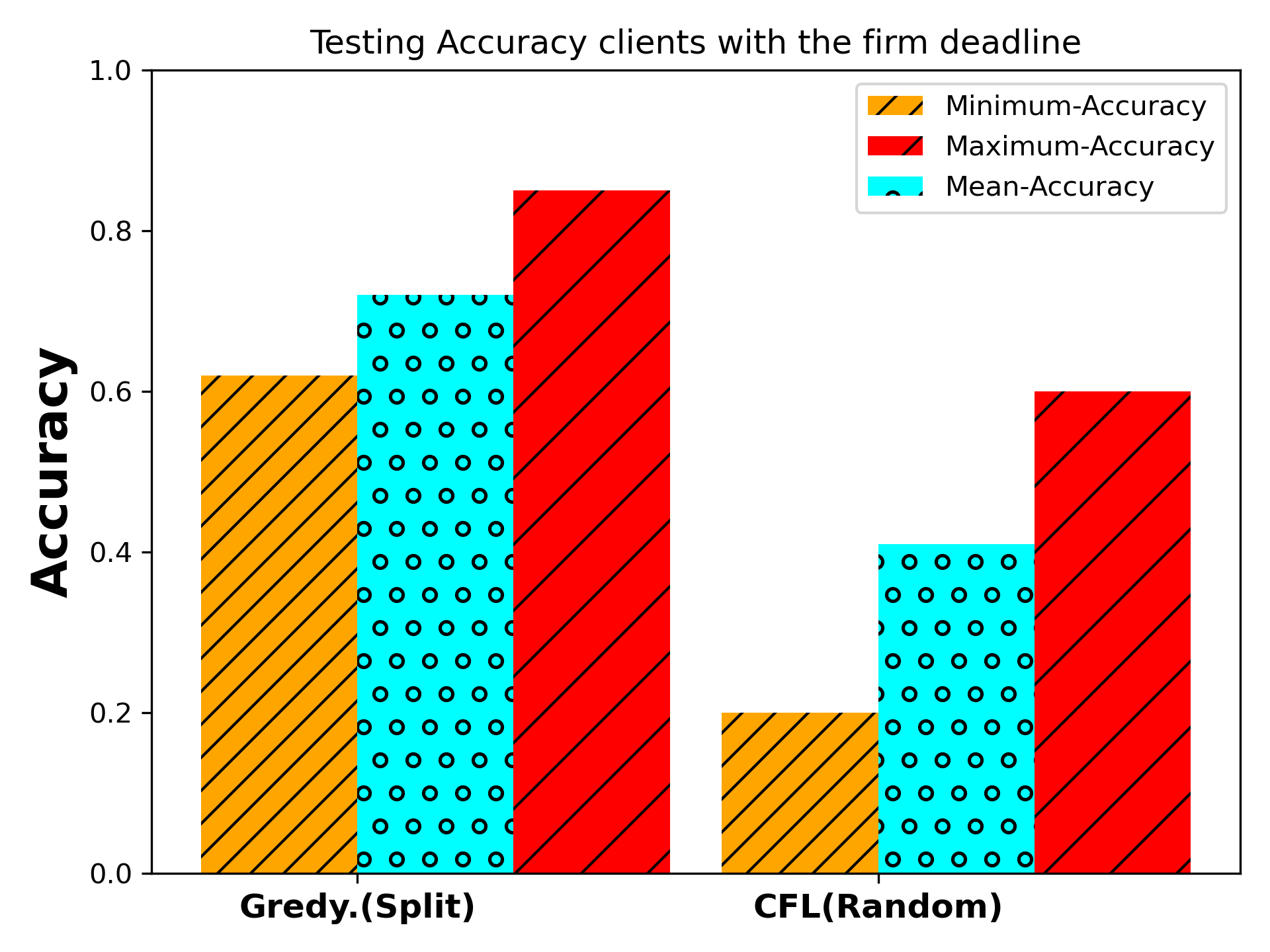}
\caption{Fair and greedy selection with split-based model aggregation.}
\label{acc_greedy_S}
\end{subfigure}
\begin{subfigure}[b]{0.23\textwidth}
\includegraphics[width=0.9\linewidth]{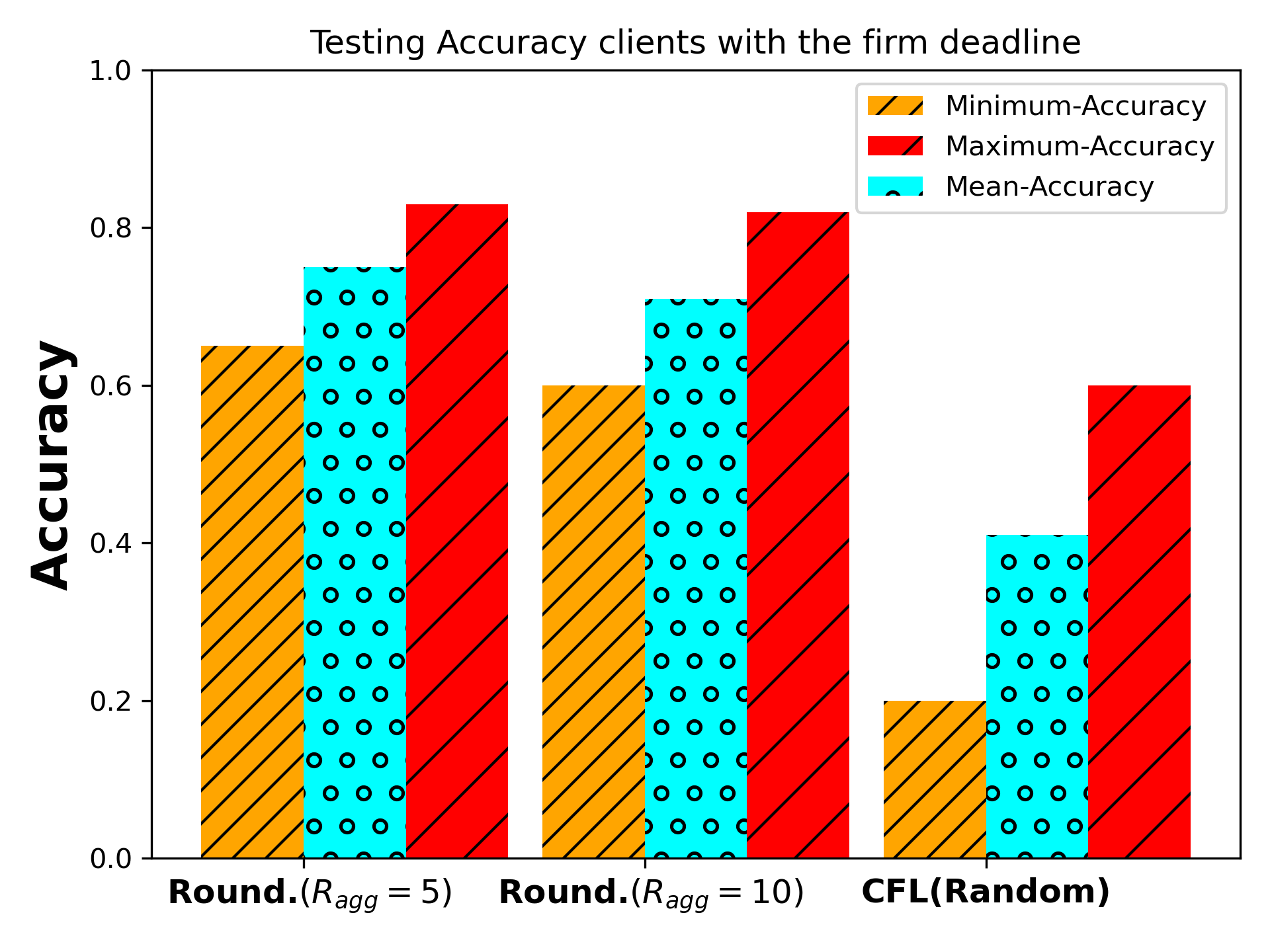}
\caption{Fair and round robin selection with round based aggregation ($R_\mathrm{agg}=10$).}
\label{acc_round_r=10}
\end{subfigure}   
\begin{subfigure}[b]{0.23\textwidth}
\includegraphics[width=0.9\linewidth]{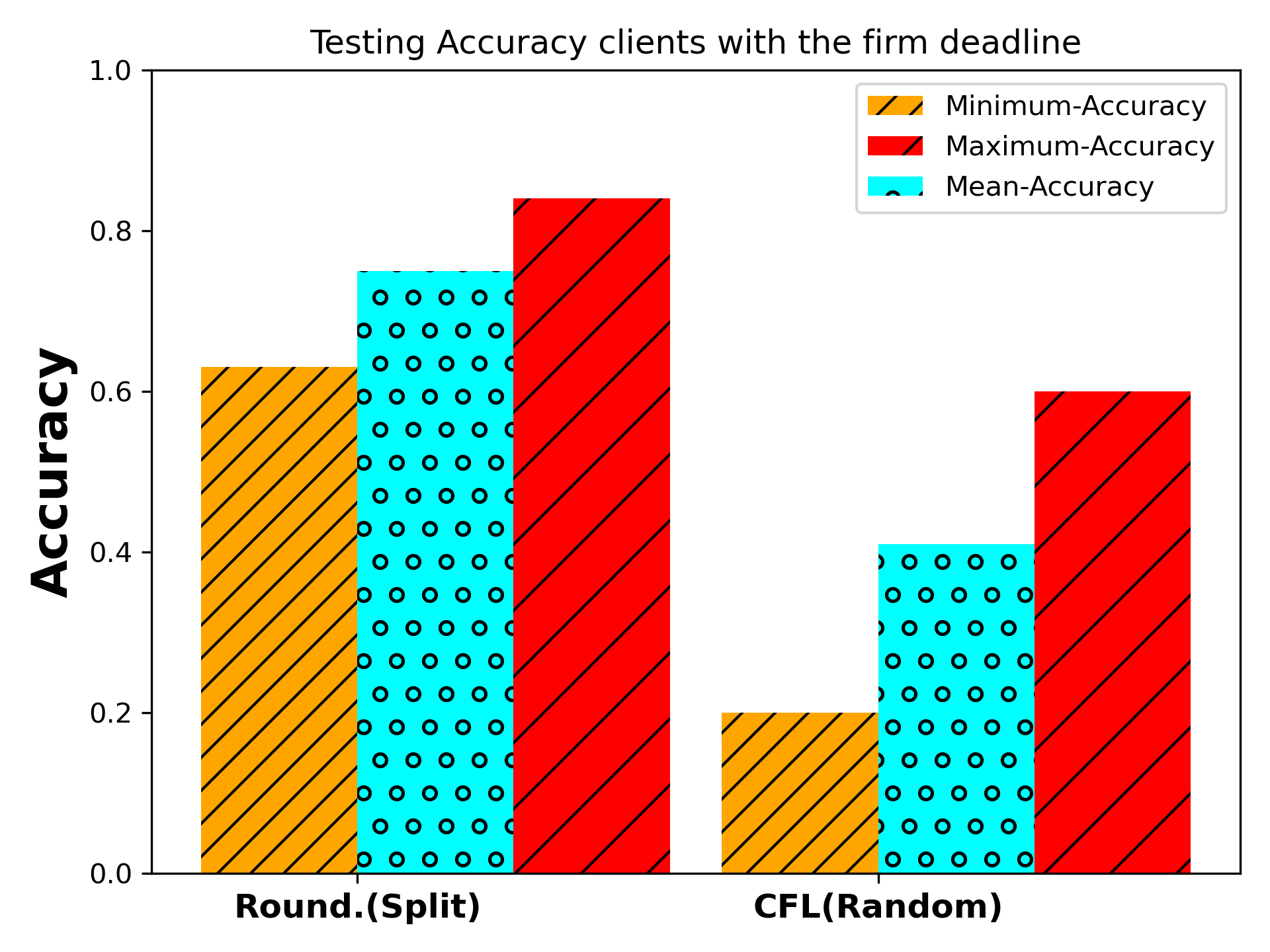}
\caption{Fair and round-robin selection with split-based model aggregation. }
\label{acc_round_S}
\end{subfigure}
\caption{Performance Evaluation of our approach regarding testing accuracy compared to baselines using FEMNIST dataset.}
\label{acc_numrical_results}
\end{figure*}
\begin{figure}[t]
\centering   
     \includegraphics[width=0.7\linewidth]{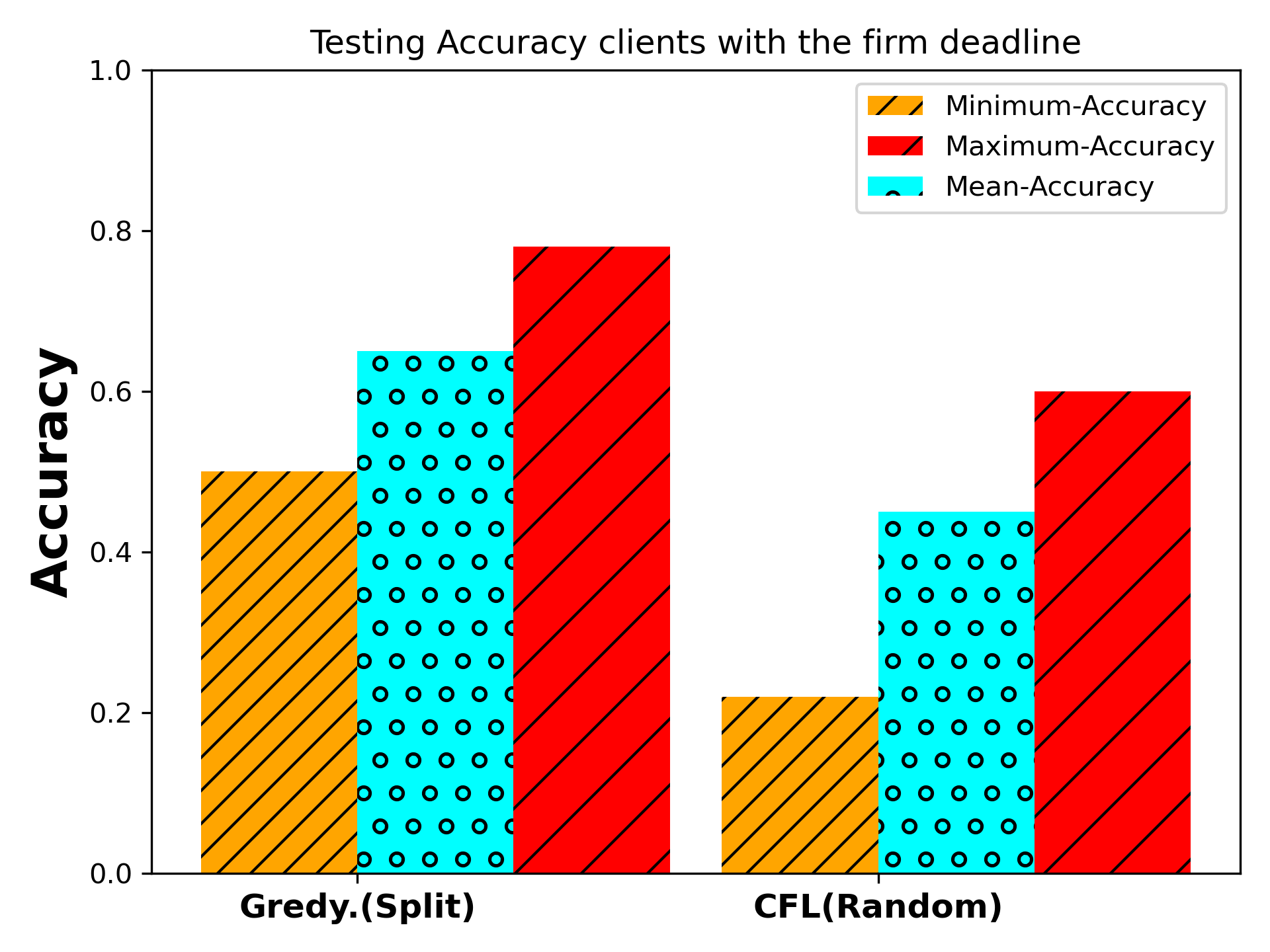}
    \caption{Testing accuracy for fair and greedy selection with split-based aggregation using CIFAR-10 dataset. }
    \label{acc_CIFAR10}
\end{figure}
 \subsection{Numerical Results}
In the setting of the experiment, we use $200$ communication rounds ($R$), $200$ clients ($N$), and $10$ epochs ($L$) in all our simulation settings.  The evaluation process for the experiments comprises two stages. Initially, experiments are conducted, and results are recorded at each global round.  Subsequently, the final models resulting from the completed training process are tested to show whether the resulting models are a better fit for all clients or not. 
\subsubsection{Importance of applying CFL in HWNs}
To validate the effectiveness of our approach in HWNs, we aim to show that it outperforms conventional FL within these networks. To this end, we select one of our previously described scenarios—specifically, ``fair and greedy selection with split-base model aggregation"—as an example of superiority and compare it to HFL, which employs random client selection and round-by-round model aggregation.

Fig. \ref{HFL with CFL} illustrates the significant improvement achieved by applying our scenario in the HWN, specifically with a two-phase client selection and two-level model aggregation. This process significantly strengthens the performance of our learning models. One can notice that there is a considerable gap, up to $140\%$, between the average performance of our proposed approach and that of HFL. This difference stems from HFL's inability to manage unbalanced and non-IID data distributions and resource constraints effectively. Through this research, we demonstrate that our approach (fair and greedy selection with split-base model aggregation)  addresses these limitations and significantly improves learning model performance. 
\subsubsection{Performance Enhancement regarding  the Clustering Speed and Convergence Rate}
The proposed approach is first evaluated in terms of the splitting process and accurate clustering using the FEMNIST dataset, as shown in Figs. \ref{greedy_r=5} to \ref{tradi_CFL}. These figures illustrate the model's performance for all scenarios for both the proposed approach and the baselines, considering testing accuracy with a confidence threshold in each round (i.e., see Fig. \ref{greedy_r=5} left) and the norm of gradients amongst all clients to highlight the accelerated convergence speed (i.e., see Fig. \ref{greedy_r=5} right). All of our scenarios are observed to attain faster convergence speed and initiate the clustering process earlier than the baselines that initiate partitioning and continue splitting through the late-stage training process. 
This represents an outstanding performance in all scenarios. In detail, the scenarios in Figs. (\ref{greedy_r=5}, \ref{greedy_r=10}, and \ref{greedy_split}, right)  finish the clients' splitting before round $25$, while the scenarios in Figs. (\ref{robin_r=5} and \ref{robin_r=10}, right) partition clients at rounds $30$ and $76$, respectively.

Regarding the stopping point, one can notice that our proposed approach reaches the stopping point for all clusters (i.e., no further split is still needed) on rounds $70$, $73$, and $125$ in all conducted scenarios. 
For instance, the fair and greedy selection with round-based model aggregation ($R_\mathrm{agg}= 5$) scenario, Fig. \ref{greedy_r=5} right, reaches the stopping point at round $73$. Also, the fair and round-robin selection with round-based model aggregation ($R_\mathrm{agg}=10$) scenario reaches this point at the $125$-\textit{th} round (Fig. \ref{robin_r=10}, right). On the other hand, we observe that as in Fig.\ref{tradi_CFL} right, the baselines still demand further rounds to reach the stopping point. 
This is mainly attributed to the random selection mechanism, which selects clients that may be previously clustered, necessitating further rounds to capture more data patterns. This implies that the random client selection does not fit the CFL in HWNs.

We further demonstrate the robustness of our proposed approach against baselines in terms of accurate clustering and splitting process using the CIFAR-10 dataset. Specifically, we pick the scenario of fair and greedy selection with split-based model aggregation as an example of the superiority of the proposed approach compared to the baselines. As illustrated in Figs. \ref{acc_CIFAR10} and \ref{CFL_CIFAR-10}, our proposed scenario outperforms the baselines, both in terms of clustering rate and convergence rate. The improved performance stems from our approach's novel strategies (i.e., two-phase client selection and two-level model aggregation), which enable more accurate client clustering and faster convergence rates. Initially, we permit all available clients to participate in the training. Subsequently, once a specific cluster approaches its stopping point, we invoke an efficient scheduling technique. By leveraging a greedy selection algorithm, clients characterized by lower latency and better resources are selected for continued training. In contrast, the baselines' performance declines due to its random client selection (i.e., no consideration for data distribution or resource availability). This non-strategic approach complicates the training process, leading to extended training time. 
\subsubsection{Learning Performance}
To showcase the learning performance of our proposed approach and the baselines, Fig. \ref{acc_numrical_results} exhibits the testing accuracy of four scenarios using the FEMNIST dataset.  Specifically, as in Figs. \ref{acc_greedy_r=5}-\ref{acc_round_S}, we simulate the testing accuracy in terms of minimum, mean, and maximum local accuracies, showing the extent to which our proposed approach can ensure the fairness of the learning performance.

For instance, a small gap in the scenario of fair and greedy selection with split-based model aggregation exists between maximum and minimum accuracy levels, showing that the resulting models align perfectly with every local data distribution (Fig. \ref{acc_greedy_S}). This improvement stems from the twofold client selection process: Firstly, we ensure fairness by incorporating all clients in the early stage of training and then applying the greedy algorithm that selects clients with less latency and better resources when a cluster reaches its stopping point. Secondly, the clients efficiently exploit their models' training before the split occurs, leading to improving their local accuracies. In contrast, the gap in the baselines between the maximum and minimum rises to significantly higher values (i.e., $62\%$) due to the random client selection for training.

In Fig. \ref{acc_round_r=10}, the scenario of fair and round-robin selection with round-based model aggregation outperforms the baselines with $R_\mathrm{agg}=5$ and $R_\mathrm{agg}=10$. The round-based model aggregation scenario with $R_\mathrm{agg}=5$ shows a significant $34\%$ mean accuracy improvement over the baselines. This mainly arises from providing the clients more time for local data training and attaining preferred accuracies before their upload to the cloud. Moreover, clients exhaust their power resources as the training period extends, necessitating varied client selections in every round. This demonstrates the significance of resource management and strategic client selection in enhancing system performance and accuracy. 

Overall, we attain substantial mean accuracy improvement in all scenarios. For instance, the fair and greedy selection with the split-based model aggregation scenario (Fig. \ref{acc_greedy_S}) surpasses the baselines by $75\%$ in the mean accuracy. This is because the $k$-\textit{th} edge server consistently employs the greedy selection for clients across global rounds by selecting the best clients with lower latency and better resources. Conversely, a significant gap exists between the maximum and minimum in the baselines due to the model's inability to contain all clients' data. This may lead to bias in selecting clients or inadequate data distribution representation, resulting in inconsistent model performance. Nonetheless, our proposed scheduling scenarios greatly outperform the baselines, delivering more consistent accuracy unaffected by data distribution or device heterogeneity in the 
 mobile edge networks within the HWN.

Similarly, we demonstrate the superior performance of our proposed approach compared to the baselines on a different dataset (CIFAR-10). Again, we focus on the scenario of fair and greedy selection with split-based model aggregation as a representative example to emphasize the advantages of our approach. Fig. \ref{acc_CIFAR10} illustrates the testing accuracy, highlighting the maximum, minimum, and mean local accuracies for both our approach and the baselines. Observations indicate that the proposed approach consistently surpasses baseline performance. This can be attributed to the scheduling strategy we employed: initially ensuring equitable participation of all available and active clients in the training phase, followed by a client selection technique (greedy selection algorithm) once a particular cluster reaches the stopping point. Overall, our approach demonstrates significant performance in both FEMNIST and CIFAR-10 datasets, surpassing baselines and evidencing robust adaptability to heterogeneous environments in HWNs.

\subsubsection{Considerable Reduction in Energy Consumption} 
\begin{figure}[t]
\centering   
     \includegraphics[width=0.7\linewidth]{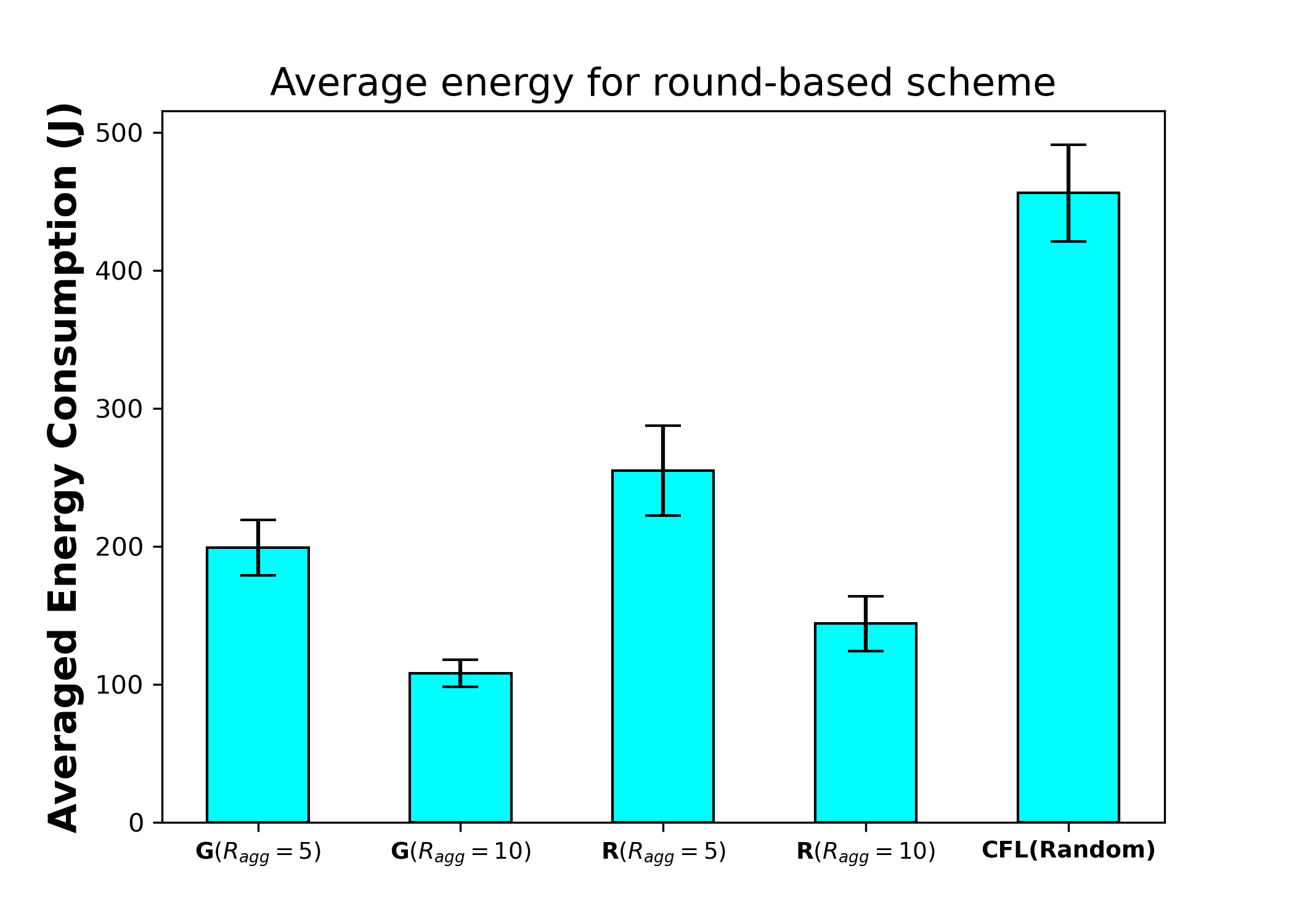}
    \caption{The round-based model aggregation with greedy and round-robin selection using the FEMNIST dataset. }
    \label{Energy_round}
\end{figure}

\begin{figure}[t]
\centering 
\includegraphics[width=0.7\linewidth]{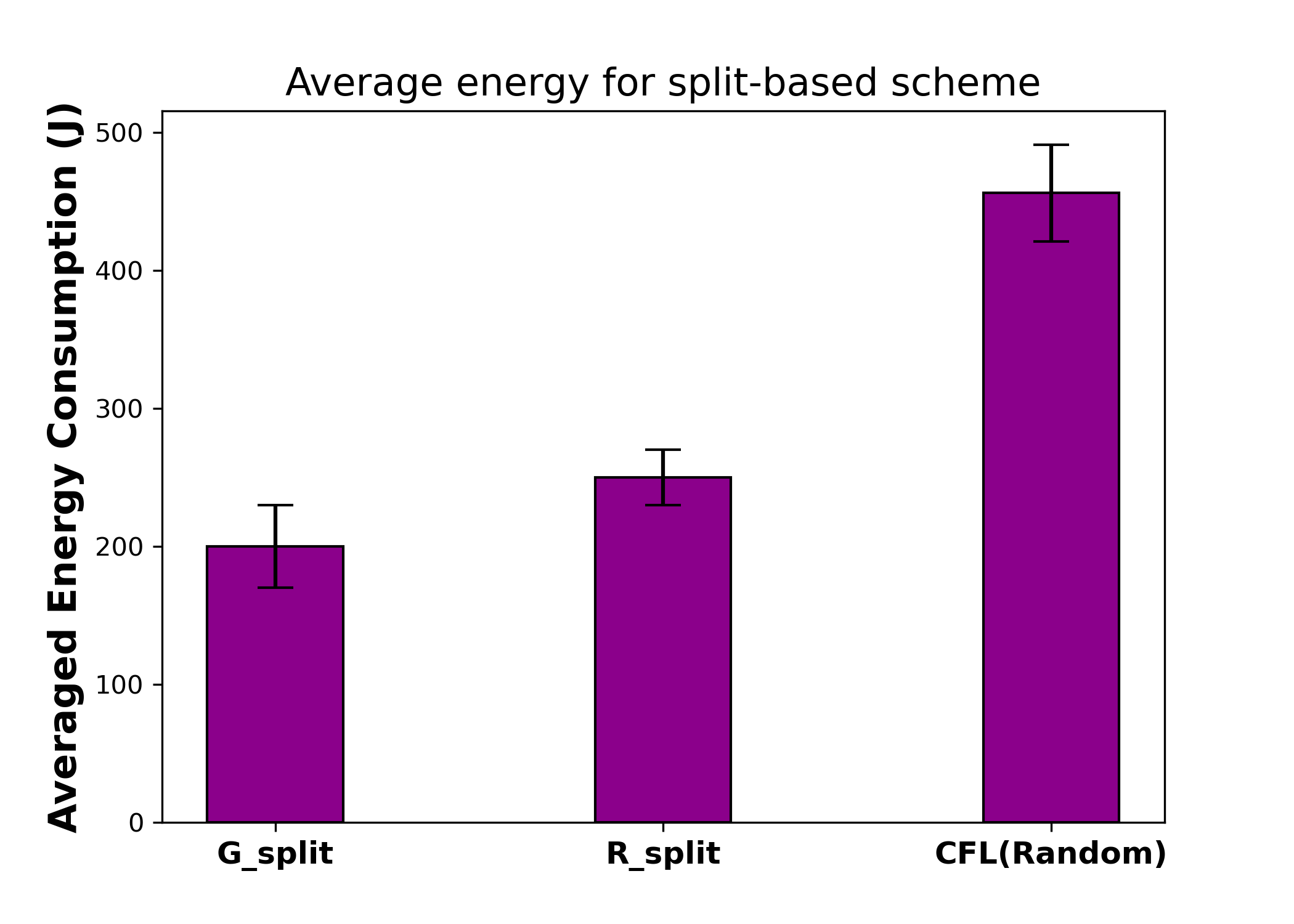}
    \caption{The split-based aggregation with greedy and round-robin selection using the FEMNIST dataset. }
    \label{Energy_split}
\end{figure}

The proposed approach exhibits a significant improvement in the averaged energy consumption compared to the baselines using the FEMNIST dataset, as shown in Figs. \ref{Energy_round} and \ref{Energy_split}. For example, Fig. \ref{Energy_round} demonstrates a significant reduction in the averaged energy consumption for the round-based model aggregation scheme at  $R_\mathrm{agg}=5$ and $R_\mathrm{agg}=10$, compared to the baselines. It is clear that for the scenario of fair and greedy selection with round-based model aggregation, the energy consumption decreases by $56\%$ when $R_\mathrm{agg}=5$ and it further reduces by $68\%$ when $R_\mathrm{agg}=10$. This is because the model aggregation occurs every five or ten rounds during the learning process, leading to reduced communication rounds frequency between the cloud and edge servers and thus saving energy consumption. Moreover, a significant reduction in energy consumption is achieved when the greedy selection algorithm is used.  This is because clients with lower latency and better resources are prioritized in training,  thus reducing energy costs. In addition, we notice that the energy consumption for the fair and greedy selection scheme proves slightly more efficient than the fair and round-robin selection when  $R_\mathrm{agg}=5$ or  $R_\mathrm{agg}=10$. This is because the greedy algorithm prioritizes clients with less latencies and better resources, while the round-robin algorithm alternates the client in each round to ensure fairness in resource distribution, inherently leading to slightly higher energy consumption.
\begin{figure}[t]
\centering 
\includegraphics[width=0.7\linewidth]{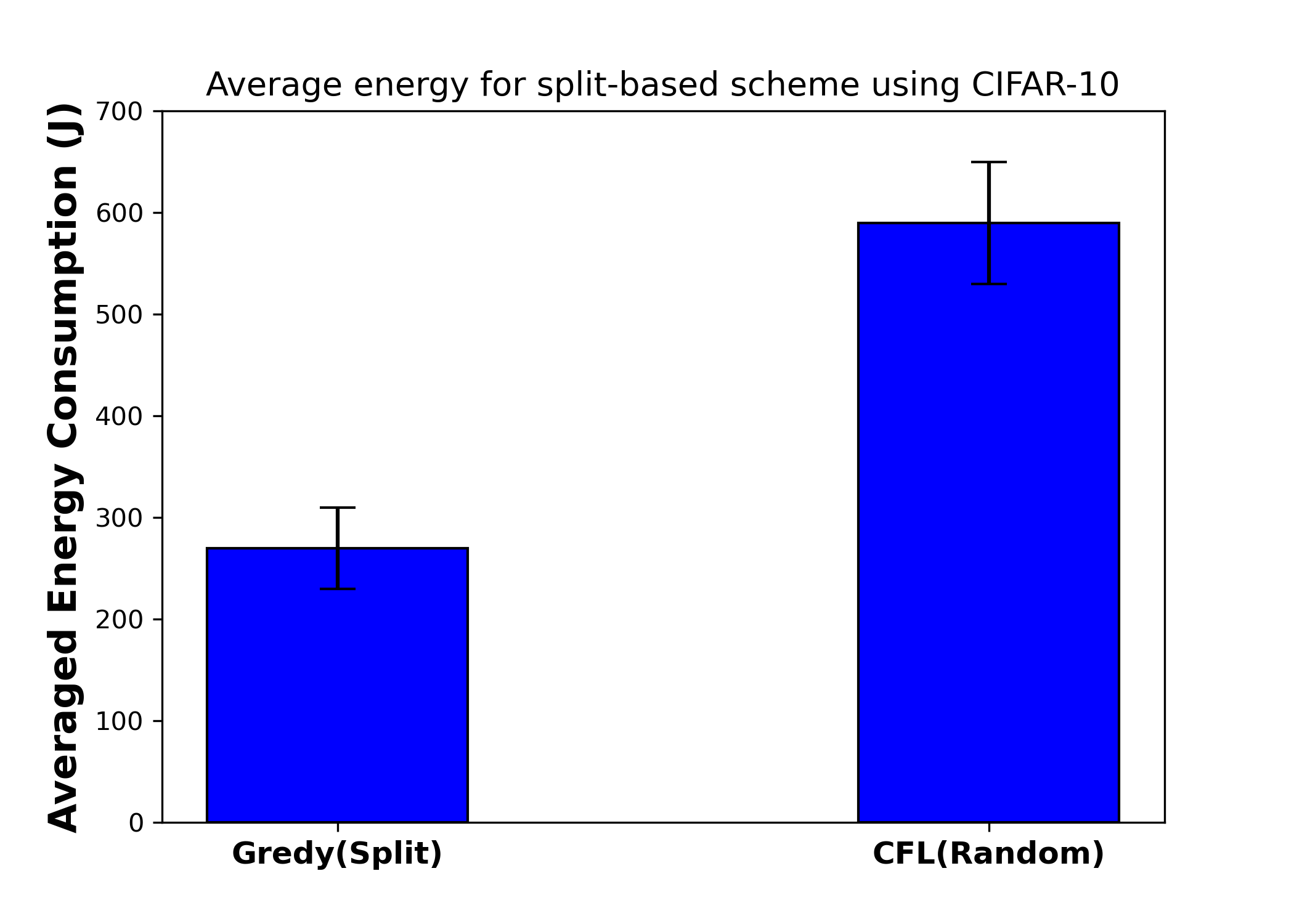}
    \caption{The split-based aggregation with greedy selection using CIFAR-10 dataset. }
    \label{Energy_split_CIFAR10}
\end{figure}

Similarly, as illustrated in Fig. \ref{Energy_split}, the split-based model aggregation scheme significantly provides energy efficiency, up to $66\%$,   compared to the baselines. This superior performance in energy consumption can be attributed to two key factors: First, our scheduling technique selects clients based on model performance (i.e., greedy selection) or fairness on resource distribution (i.e., round-robin selection). Second, the split is only performed when conditions (\ref{cond1}) and (\ref{cond2}) are met during training, which triggers model aggregation in the cloud. Our findings confirm that our proposed approach outperforms the baseline techniques, exhibiting superior performance in both system model effectiveness and resource optimization. 

To evaluate the proposed approach in terms of the averaged energy consumption using the  CIFAR-10 dataset, refer to Fig. \ref{Energy_split_CIFAR10}. This figure highlights a substantial reduction in the averaged energy consumption for the scenario of fair and greedy selection with a split-based model aggregation compared to the baselines. Notably, our proposed approach reduces energy consumption by $54\%$ compared to baselines. This significant reduction can be attributed to our scheduling technique and model aggregation schemes that we employ in HWNs. In particular, once a cluster reaches the stopping point, the proposed approach uses a greedy selection where clients exhibiting less latency and better resources are selected for further training. In addition, the proposed approach leverages split-based model aggregation, activating model aggregation only when one split is performed on the edge servers. In general, our approach consistently outperforms baselines in averaged energy consumption across different datasets (i.e., FEMNIST and CIFAR-10). This emphasizes the robustness and flexibility of the proposed approach within HWNs.
\subsubsection{Lessons Learned}
\begin{itemize}
    \item The proposed approach improves the convergence speed compared to the baselines due to the fairness implemented among clients before the clusters reach their stopping points.
    \item We avoid creating biased models, which can result from frequently selecting the same clients during training, by designing tailored models for each client, which are a better fit with diverse data distributions.
    \item Providing collaborative learning across mobile edge networks is achieved by using HWNs, which enable knowledge sharing through tailored models that accurately accommodate the diverse data patterns from different mobile edge networks.
    \item Resource consumption is significantly reduced due to two key factors: (i) the model aggregation only occurs either at the aggregation rounds or when the edge servers perform at least one split, and  (ii)  training is initiated by a single client, as in greedy selection, or in a cyclic manner, as in round-robin selection once each cluster reaches its stopping point.
    \item Overall, the proposed approach demonstrates superior performance over the baselines in terms of creating highly customized models, accelerating the convergence rate, and reducing both the training time and energy expenditure.
\end{itemize} 
\section{Conclusion}
\label{conclusion}
This work introduced a novel framework comprising two-phase client selection and two-level model aggregation for CFL in HWN. Our proposed approach reduced the training time and resource consumption while ensuring satisfactory client accuracy and accelerating the convergence rate. It also ensured fairness among clients by offering them equal opportunities to participate in the training and designing a tailored model for each client. We formulated an optimization problem to identify optimal client selection and model aggregation schemes. This process enables the HWNs to minimize training costs and enhance the speed of convergence in the presence of unbalanced and non-IID data distribution and the diversity of clients. We conducted comprehensive simulation experiments using realistic federated datasets (FEMNIST and CIFAR-10).
The numerical results showed the superior performance of our proposed approach over baseline techniques in terms of computation and communication time, convergence speed, and resource consumption, all while providing satisfactory performance. In our future works,  we aim to explore learning in dynamic environments where data is non-static and continually evolving.
\section*{Acknowledgement}
{
This publication was made possible by TUBITAK-QNRF joint Funding Program (Tubitak-QNRF 4th Cycle grant \# AICC04-0812-210017) from the Qatar National Research Fund (a member of Qatar Foundation). The findings herein reflect the work, and are solely the responsibility of the authors.}
\bibliographystyle{IEEEtran}
\bibliography{Ref}
\begin{IEEEbiography}[{\includegraphics[width=1.1in,height=1.22in,clip]{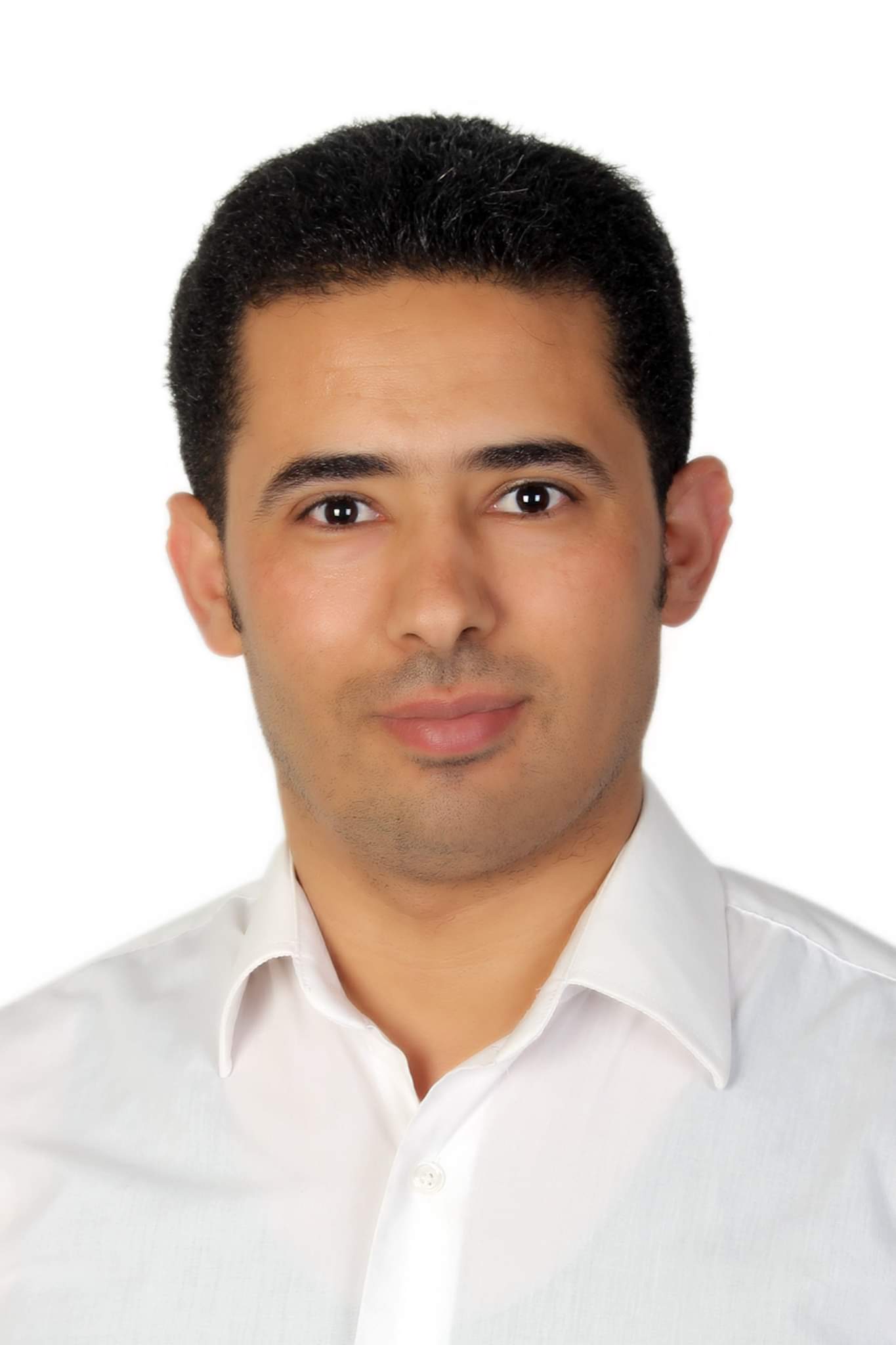}}]  
{Moqbel Hamood} received his B.Sc. degree in Electrical Engineering from Mutah University, Jordan, in 2012 and his M.Sc. degree in Wireless Communications from Jordan University of Science and Technology (JUST), Irbid, Jordan, in 2018. He is currently pursuing his Ph.D. at the Smart Communication Networks \& Systems Lab at Hamad Bin Khalifa University, Doha, Qatar. His research interests include federated learning over wireless networks.
\end{IEEEbiography}
\begin{IEEEbiography}[{\includegraphics[width=1.1in,height=1.22in,clip]{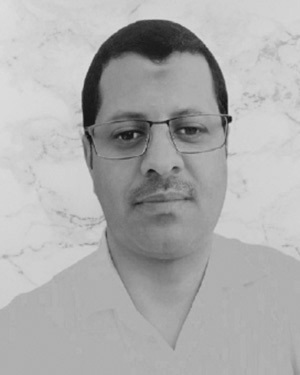}}]  
{Abdullatif Albaseer (Member, IEEE)} received his M.Sc. degree (with honors) in computer networks from King Fahd University of Petroleum and Minerals, Dhahran, Saudi Arabia, in 2017, and his Ph.D. degree in computer science and engineering from Hamad Bin Khalifa University, Doha, Qatar, in 2022. He is currently a Postdoctoral Research Fellow with the Smart Cities and IoT Lab at Hamad Bin Khalifa University. He was a Visiting Scholar at Texas A \& M University in Qatar from 2022 to 2023.
Dr. Albaseer has authored and co-authored over thirty journal and conference papers, primarily in IEEE Transactions. He also holds six US patents in the area of wireless network edge technologies. His current research interests include AI for Networking, AI for Cybersecurity, and the application of Large Language Models (LLMs) in both fields.
Dr. Albaseer has served as a chair and organizing committee member for international conferences. He is a reviewer for numerous prestigious IEEE journals and conferences. He has also presented at various international conferences and participated in numerous academic and professional development activities.
\end{IEEEbiography}
\begin{IEEEbiography}[{\includegraphics[width=1.1in,height=1.22in,clip]{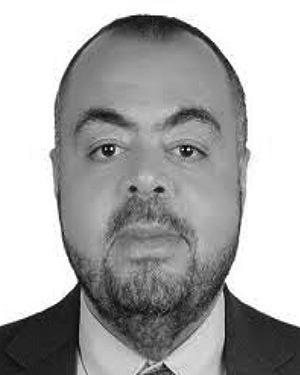}}]  
{Mohamed Abdallah (Senior Member, IEEE)} received the B.Sc. degree from Cairo University, in 1996, and the M.Sc. and Ph.D. degrees from the University of Maryland at College Park in 2001 and 2006, respectively. From 2006 to 2016, he held academic and research positions with Cairo University and Texas A\&M University at Qatar. He is currently a Founding Faculty Member with the rank of an Associate Professor with the College of Science and Engineering, Hamad Bin Khalifa University. He has published more than 150 journals and conferences and four book chapters, and co-invented four patents. His current research interests include wireless networks, wireless security, smart grids, optical wireless communication, and blockchain applications for emerging networks. He was a recipient of the Research Fellow Excellence Award at Texas A\&M University at Qatar in 2016, the Best Paper Award in multiple IEEE Conferences, including the IEEE BlackSeaCom 2019, the IEEE First Workshop on Smart Grid and Renewable Energym in 2015, and the Nortel Networks Industrial Fellowship for five consecutive years from 1999 to 2003. His professional activities include an Associate Editor of the IEEE Transactions on Communications and the IEEE Open Access Journal of Communications, the Track Co-Chair of the IEEE VTC Fall 2019 Conference, the Technical Program Chair of the 10th International Conference on Cognitive Radio Oriented Wireless Networks, and a Technical Program Committee Member of several major IEEE conferences.
\end{IEEEbiography}
\begin{IEEEbiography}[{\includegraphics[width=1.1in,height=1.22in,clip]{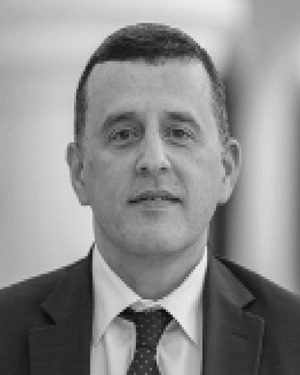}}]  
{Ala Al-Fuqaha (Senior Member, IEEE)} received the Ph.D. degree in computer engineering and networking from the University of Missouri-Kansas City, Kansas City, MO, USA, in 2004. He is currently a Professor with Hamad Bin Khalifa University. His research interests include the use of machine learning in general and deep learning in particular in support of the data-driven and self-driven management of large-scale deployments of IoT and smart city infrastructure and services, wireless vehicular networks (VANETs), cooperation, and spectrum access etiquette in cognitive radio networks, and management and planning of software defined networks. He serves on editorial boards of multiple journals, including IEEE Communications Letter and IEEE Network Magazine. He also served as the Chair, the Co-Chair, and a Technical Program Committee Member of multiple international conferences, including IEEE VTC, IEEE Globecom, IEEE ICC, and IWCMC. He is an ABET Program Evaluator.
\end{IEEEbiography}
\begin{IEEEbiography}[{\includegraphics[width=1.1in,height=1.22in,clip]{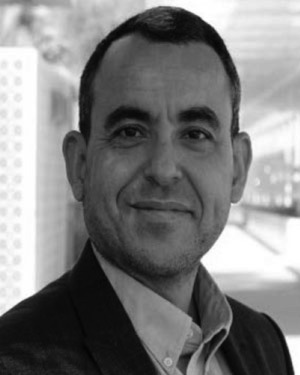}}]  
{Amr Mohamed (Senior Member, IEEE)} received the M.S. and Ph.D. degrees in electrical and computer engineering from the University of British Columbia, Vancouver, Canada, in 2001, and 2006, respectively. He has worked as an advisory IT specialist in IBM Innovation Center, Vancouver, from 1998 to 2007, taking a leadership role in systems development for vertical industries. He is currently a Professor with the College of Engineering, Qatar University and the Director of the Cisco Regional Academy. He has over 25 years of experience in wireless networking research and industrial systems development. He has authored or coauthored over 200 refereed journal and conference papers, textbooks, and book chapters in reputable international journals, and conferences. His research interests include wireless networking and edge computing for IoT applications. He holds 3 awards from IBM Canada for his achievements and leadership, and the four best paper awards from IEEE conferences. He is serving as a technical editor for two international journals and has served as a technical program committee co-chair for many IEEE conferences and workshops.
\end{IEEEbiography}
\end{document}